\documentclass[graybox]{svmult}

\usepackage{type1cm}        % activate if the above 3 fonts are
\usepackage{makeidx}         % allows index generation
\usepackage{graphicx}        % standard LaTeX graphics tool
\usepackage{multicol}        % used for the two-column index
\usepackage[bottom]{footmisc}% places footnotes at page bottom

\usepackage{newtxtext}       % 
\usepackage{newtxmath}       % selects Times Roman as basic font

\usepackage{xcolor} % colored tables

\usepackage{graphicx}% Include figure files
\usepackage{dcolumn}% Align table columns on decimal point
\usepackage{bm}% bold math
\usepackage[flushleft]{threeparttable} % table footnote

\usepackage{dsfont}
\usepackage{color,psfrag}
\usepackage{mathtools,amscd}

\usepackage{soul}
\usepackage{xcolor} % colored tables

\usepackage{ifpdf}
    \ifpdf
\usepackage{tikz}
\usetikzlibrary{arrows,chains,matrix,positioning,scopes, fit, calc}
\usetikzlibrary{cd}
\usepackage{chemfig}
\fi
\usepackage{hyperref}

\hypersetup{
	colorlinks,
	linkcolor=blue,
	citecolor=blue, 
	filecolor=black,
	urlcolor=blue}

\newcommand{\mat}[1]{\mathbf{#1}} % matrices in bold

\newcommand{\tran}{^{\mathstrut\scriptscriptstyle\top}} 
\makeindex             % used for the subject index
\begin{document}

\title*{Construction of Machine Learned Force Fields with Quantum Chemical Accuracy: Applications and Chemical Insights}
\author{Huziel E. Sauceda$^{a}$, Igor Poltavsky$^{b}$, Stefan Chmiela$^{c}$, Klaus-Robert M\"uller$^{c,d,e}$ and Alexandre Tkatchenko$^{b}$}
\institute{Huziel E. Sauceda, \email{sauceda.he@gmail.com}
\and Klaus-Robert M\"uller, \email{klaus-robert.mueller@tu-berlin.de}
\and Alexandre Tkatchenko, \email{alexandre.tkatchenko@uni.lu} 
\at $^a$Fritz-Haber-Institut der Max-Planck-Gesellschaft, 14195 Berlin, Germany.
$^b$Physics and Materials Science Research Unit, University of Luxembourg, L-1511 Luxembourg.
$^c$Machine Learning Group, Technische Universit\"at Berlin, 10587 Berlin, Germany.
$^d$Department of Brain and Cognitive Engineering, Korea University, Anam-dong, Seongbuk-gu, Seoul 02841, Republic of Korea. 
$^e$Max Planck Institute for Informatics, Stuhlsatzenhausweg, 66123 Saarbr\"ucken, Germany.}
\maketitle

\abstract{Highly accurate force fields are a mandatory requirement to generate predictive simulations. 
Here we present the path for the construction of machine learned molecular force fields by discussing the hierarchical pathway from generating the dataset of reference calculations to the construction of the machine learning model, and the validation of the physics generated by the model. 
We will use the the symmetrized gradient-domain machine learning (sGDML) framework due to its ability to reconstruct complex high-dimensional potential-energy surfaces (PES) with high precision even when using just a few hundreds of molecular conformations for training. 
The data efficiency of the sGDML model allows using reference atomic forces computed with high-level wavefunction-based approaches, such as the \textit{gold standard} coupled cluster method with single, double, and perturbative triple excitations (CCSD(T)). 
We demonstrate that the flexible nature of the sGDML framework captures local and non-local electronic interactions (e.g. H-bonding, lone pairs, steric repulsion, changes in hybridization states (e.g. $sp^2 \rightleftharpoons sp^3$), $n\to\pi^*$ interactions, and proton transfer) without imposing any restriction on the nature of interatomic potentials. 
The analysis of sGDML models trained for different molecular structures at different levels of theory (e.g. density functional theory and CCSD(T)) provides empirical evidence that a higher level of theory generates a smoother PES. 
Additionally, a careful analysis of molecular dynamics simulations yields new qualitative insights into dynamics and vibrational spectroscopy of small molecules close to spectroscopic accuracy.}

\section{Introduction} 
\label{sec:intro}
\textit{In silico} studies of molecular systems and materials constitute one of the most important tools in physics, biology, materials science, and chemistry due to their great contributions in understanding systems ranging from small molecules (e.g. few atoms) up to large proteins and amorphous materials, guiding the exploration and the discovery of new materials and drugs. 
This requires the construction of physical models that faithfully describe interatomic interactions, and quantum mechanics (QM) is the pertinent methodology to engage such monumental task.
Nevertheless, using the full machinery of QM (e.g. Dirac equation~\cite{DIRAC18} and Quantum Electrodynamics~\cite{CCQED_Gold_PRL2017}) would lead not far from simulations of diatomic molecules.

To overcome this limitation, for most of the problems of interest, one can approximately describe a molecular system by the more tractable non-relativistic time-independent Sch\"odinger equation.

Additionally, one often decouples nuclear and electronic degrees of freedom by employing the Born-Oppenheimer (BO) approximation.
%, $\Phi=\Phi_{nucl}\Psi_{elec}$. 
This makes predictive simulations of molecular properties and thermodynamic functions possible by representing a $N$-atoms system by the global potential energy surface (PES) $V_{BO}(\mathbf{x})$ where $\mathbf{x}=\{\mathbf{r}_1,\mathbf{r}_2,...,\mathbf{r}_N\}$ and $\mathbf{r}_i$ the $i$th nuclear Cartesian coordinates. 
$V_{BO}(\mathbf{x})$ is defined as the sum of the total electrostatic nuclear repulsion energy $\sum_{i,j>i}{Z_i Z_j r_{ij}^{-1}}$ and the electronic energy $\mathcal{E}_{\text{elec}}$ solution of the electronic Schr\"odinger equation $\mathcal{H}_{\text{elec}}\Psi=\mathcal{E}_{\text{elec}}\Psi$ for a given set of nuclear coordinates $\mathbf{x}$.
Therefore, $V_{BO}$ contains all the information necessary to describe nuclear dynamics of the molecular system since all electronic quantum interactions are encoded in it via $\mathcal{E}_{\text{elec}}$ within the BO approximation.
A systematic partitioning of this energy could potentially help to gain further insights into the physics and chemistry of the system,
nevertheless, in practice it is not known how to exactly expand the $V_{BO}$ in different energetic contributions such as hydrogen bonding, electrostatics, dispersion interactions or other electronic effects.
Furthermore, any attempt in separating the PES in terms of known analytic forms or empirically derived interactions will always result in biasing the final model which limits its possible accuracy and may introduce non-physical artifacts.
% possibly introducing non-physical interactions or artifacts. 
%
Therefore, the intricate form of $V_{BO}$ resulting from an interplay between different quantum phenomena when solving the Schr\"odinger equation should be preserved.

%MD intro
In order to extract the dynamical properties and thermodynamics of molecular systems, the $V_{BO}$ has to be sampled according to a thermodynamic ensemble (e.g. NVE, NVT, $\mu$VT, etc.) depending on the property being computed. 
The two most popular techniques are Monte Carlo sampling and molecular dynamics simulations (MD). In particular, MD constitutes the fundamental pillar of contemporary science by allowing remarkable advances and offering unprecedented insights into complex chemical and biological systems.
However, sampling the $V_{BO}$ using this technique in any of its flavors (e.g. Langevin or Verlet-velocity propagator) to obtain converged mechanical and thermodynamical properties often requires millions integration steps, meaning that the Schr\"odinger equation $\mathcal{H}_{\text{elec}}\Psi=\mathcal{E}_{\text{elec}}\Psi$ has to be solved and $-\textbf{F}=\langle \Psi^*|\partial \mathcal{H} /\partial \textbf{x}|\Psi \rangle$ evaluated a similar amount of times~\cite{TuckermanBook}. 
Such direct \textit{ab initio} molecular dynamics (AIMD) simulations, where the quantum-mechanical energies and forces are computed on-the-fly for molecular configurations at every time step, are known to generate highly accurate but computationally very costly predictions. 
In practice, most of the works in AIMD use density functional theory (DFT) to approximate the solution of the Schr\"odinger equation for a system of electrons and nuclei. 
Unfortunately, in some cases different exchange-correlation functionals yield contrasting results for molecular properties~\cite{Koch-HolthausenBook} and it is not clear how to systematically improve their performance. 
Alternatively, wave-function based methods that account for electron correlation (e.g. post-Hartree--Fock methods) offer a systematically improvable framework but they are rarely used in AIMD simulations due to the steep increase in the required computational resources. 
For example, a nanosecond-long AIMD simulation for a single ethanol molecule using CCSD(T) method would demand approximately a million CPU years on modern hardware. 

%Intro FFs
It is clear that AIMD is not an affordable route to pursue predictive simulations for most of the systems of interest. 
An alternative is to roughly approximate the $V_{BO}$ by creating handcrafted interatomic and physically inspired potentials with parameters fitted to experimental data or quantum-mechanical calculations. 
This has been a common practice since the early works on molecular dynamics~\cite{Alder1959,Rahman1964,Verlet1967,Rahman1971}. 
The complexity of creating reliable interatomic potentials using prior physical knowledge led to the development of dedicated force fields (FFs) for different chemical systems, a successful approach as highlighted by the 2013 Nobel Prize in Chemistry.
Examples are the TIP\textit{n}P FFs for water~\cite{TIP4P1983,TIP5P2000}, Tersoff potential for covalent materials~\cite{Tersoff1988}, polarizable FFs~\cite{PolarFF2006}, tight-binding potentials for semiconductors and metals~\cite{EAM1984}. 
This also includes a plethora of biomolecular FFs such as AMBER, MMFF, CHARMM, and GROMOS; FFs that often give reliable results for protein folding under ambient conditions~\cite{AMBER1981,CHARMM1983,MMFF94,GROMOS2005}.
The wide variety of available interatomic potentials 
%pictures the vast amount of fundamentally different atomic interactions
highlights the fact that handcrafting a FF capable of describing different types of interactions (metallic bonding, covalent chemistry, hydrogen bonding, non-covalent interactions, etc.) in a unified and seamless fashion is a complex challenge. 
Furthermore, it is widely recognized that even dedicated molecular mechanic FFs can not generate quantitative predictions from MD simulations due to their lack of accuracy. These increasingly pressing issues hinder truly predictive modeling, but at the same time encourage the development of more accurate and efficient methodologies. 

% ML-FFs
One of the possible pathways is the employment of machine learning (ML) methods for the reconstruction of the PES function.
Machine learned force fields (ML-FFs) exploit the correlation encoded in molecular datasets generated from AIMD trajectories (or any other sampling methodology) to reconstruct the underlying PES without imposing any particular explicit analytic form for the interatomic interactions. 
Furthermore, machine learning is based on rigorous statistical learning theory~\cite{SLT1,SLT2}, providing a powerful and general framework for FF learning. 
ML approaches can reconstruct complex high-dimensional objects with arbitrary precision given sufficient amount of data samples (e.g. molecular energies and atomic forces) for training.
The accurate learning of $V_{BO}$ is not a trivial task and it has driven a vast amount of work such as data sampling~\cite{DeVita2015,Shapeev2017,dral2017structure,noe2018,noe2018b}, molecular representations~\cite{Rupp2012,Bartok2010,Hansen2013,Bartok2015_GAP,Rupp2015,Ceriotti2016,artrith2017efficient,Ceriotti2017,Glielmo2017,yao2017many,faber2017prediction,eickenberg2018solid,glielmo2018efficient,Grisafi2018,tang2018atomistic,pronobis2018many,FCHL2018}, neural networks architecture development~\cite{Behler2007,Behler2012,Behler2016,Gastegger2017,dtnn,SchNet2018,SchNetNIPS2017,ryczko2018convolutional,zhang2018deep}, inference methods~\cite{Bartok2013,Montavon2013a,Ramprasad2015,Brockherde2017,huan2017universal,Tristan2018,lubbers2018hierarchical,kanamori2018exploring,hy2018predicting,Smith2017,Clementi2018,winter2019,ResponseFieldAnatole2018,gdml,sgdml,sGDMLjcp} and explanation methods~\cite{innvestigate,meila2018,Lapuschkin2019,ExplainableAIBook}. 
A crucial contribution to the further development and understanding of the field is the releasing of ready-to-use software as well as molecular datasets which guaranties the reproducibility of published results~\cite{sGDMLsoftware2019,yao2018tensormol,schnetpack2018,meila2018}.
In terms of the performance, the computational cost of evaluating ML-FFs lies in between molecular mechanic FFs and \textit{ab initio} calculations. 
In particular, the sGDML framework~\cite{gdml,sgdml} is 5-10 orders of magnitude faster than \textit{ab initio} calculations and 2-3 orders of magnitude slower than molecular mechanic FFs.\footnote{It is important to notice that while the scaling of the performance in ML-FFs depends \textit{only} on the number of atoms, while in the case of \textit{ab initio} quantum chemical calculations their performance depends on the level of theory and on the size of the basis used to approximate the wave-function and the number of electrons.}
A precise number depends  on the molecular system under study. As a reference, the sGDML model can be up to 10$^7$ and 10$^9$ times faster than CCSD(T)/cc-pVTZ level of theory for a single point calculation of malondialdehyde and aspirin, respectively~\cite{sGDMLsoftware2019}, preserving the same accuracy. This allows the use of these ML-FFs for performing long-time MD simulations and exploring different molecular properties on the CCSD(T)-level of accuracy.   

The PES reconstruction problem can be approached from two different but in principle equivalent ways\footnote{To the best knowledge of the authors up to this day these are the only two ways have been used in the PES reconstruction problem.}, by learning directly the scalar function $V_{BO}$ or by first reconstructing the gradient field associated to the PES, $\nabla V_{BO}$, and then recover the PES by analytic integration. 
These two types of ML models are called energy $\hat{f}_E$ and force $\mat{\hat{f}_F}$ models, respectively\footnote{The symbol $\hat{f}$ will be reserved to represent the predictor function of the machine learning model.}.
The two most established methodologies to create such models are Neural Networks (NN)~\cite{Behler2007,Behler2007b,Behler2011,Behler2011a,Behler2012,Behler2016,Gastegger2017,SchNetNIPS2017,SchNet2018} and kernel methods~\cite{Bartok2013,Bartok2015_GAP,Ramprasad2015,Rupp2015,DeVita2015,eickenberg2018solid,Shapeev2017,Glielmo2017,noe2018,ResponseFieldAnatole2018,gdml,sgdml}. 
%In both approaches energies and/or forces can be used for training.
%
An energy model, $\hat{f}_E$, can be based on NNs or kernel methods and trained on energies or using a combination of energies and forces~\cite{Behler2007,Gastegger2017,dtnn,SchNetNIPS2017,SchNet2018,Bartok2013,Bartok2015_GAP,Ramprasad2015,eickenberg2018solid,Shapeev2017}.
The associated FF to $\hat{f}_E$ is generated by analytic differentiation, $\mat{\hat{f}}_{\mat{F}\gets E} = - \nabla \hat{f}_E$, which introduces some disadvantages to be discussed further in the text.
In the case of force models $\mat{\hat{f}_F}$, they could also be constructed using NNs but the problem is that to recover the underlying PES requires the analytic integration of the vector field predictor. 
This immensely limits their applicability since without an appropriate integration scheme they will not be able to recover the PES.
A more common way to generate force models is using kernel methods~\cite{DeVita2015,Glielmo2017,ResponseFieldAnatole2018,gdml} usually trained directly in the gradient domain.

Contrary to the case of NNs based force models, kernels methods offer a much more flexible framework to conveniently define its analytic form, this is done by utilizing the robust framework of Gaussian processes which allows the incorporation of prior physical knowledge.
Therefore, recovering the underlying PES $\hat{f}_{E \gets \mat{F}}$ can be easily done by imposing that the mathematical formulation of $\mat{\hat{f}_F}$ to be analytically integrable and consequently it will, by definition, encode the fundamental physical law of energy conservation~\cite{gdml}.

In the limit of an infinite amount of data, energy and force models should converge to the same prediction error. 
Nevertheless, when dealing with finite or restricted amounts of data these two models do present very different performances. 
Some of the fundamental advantages of using force models instead of energy models are:
(i) Learning in the gradient domain yield smoother PESs, (ii) training exclusively on forces generates more accurate models than training using energies or a combination of both~\cite{sGDMLjcp,SchNet2018,sGDMLsoftware2019}, (iii) obtaining energies by analytical integration of force models tends give better behaved predictions as a result of the integral operator, this is in contrast with forces generated out of energy models by the gradient operator~\cite{sGDMLjcp}, and (iv) force models are more data efficient~\cite{gdml,sgdml}. 
It is important to highlight that the data efficiency of force models arises not only because the greater amount of information in each force sample (3$N$ components, where $N$ is the number of atoms), but also because each entry of the force vector is orthogonal to the rest\footnote{The components of the force vector are orthogonal in $\mathbb{R}^{3N}$, space where the function is defined.}, therefore providing a complete linearized description of its immediate local neighborhood~\cite{LearnLinNIPS2002}.
Continuing with the discussion of data efficiency, there is only a handful of models that fulfill this requirement. 
Even though formally both NN and kernel-based methods can achieve any desired accuracy, the realm of scarce data belongs to kernel models.\footnote{The reason of such difference between NNs and kernel models is that, while kernels rely on feature engineering (i.e. handcrafted descriptors), NNs represent an end-to-end formalism to describe the data. This means that NNs require more data to infer the representation that optimally describes the system.}
This is the case, for example, when the system under study requires to be described by a highly accurate reference method and it is only possible to compute a couple of hundreds of data points, as would be the case of some of the aminoacids or large molecules. 
Such better reconstruction efficiency of kernel methods is due to their greater use of prior information, offering a unique and well-defined solution~\cite{gdml}.

%sGDML intro

Here, we will focus on the \textit{\textbf{s}}ymmetric \textit{\textbf{G}}radient \textit{\textbf{D}}omain \textit{\textbf{M}}achine \textit{\textbf{L}}earning (sGDML) FF~\cite{sgdml}. The sGDML is a kernel-based ML model which directly learns forces since it is trained explicitly in the gradient domain of $V_{BO}$. 
The principal feature of this model is that it was mathematically conceived as a analytically-integrable curl-free framework. The energy conservation law is explicitly encoded into the model. Therefore, once the sGDML-FF~$\mat{\hat{f}_F}$ is trained, the potential energy function $\hat{f}_{E\gets \mat{F}}$ is also available. %
It is worth highlighting that only forces are used for training given that there is empirical evidence that a loss function that combines energies and forces causes a degradation in the force prediction~\cite{sGDMLsoftware2019,SchNet2018,ChmielaThesis}. 
The second fundamental property of the model is that the complexity of the reconstruction process is reduced through the explicit incorporation of molecular symmetries (i.e. rigid and fluxional). These permutational symmetries are automatically extracted from the reference dataset via a multi-partite procedure~\cite{MultyPartiteMatchNIPS2013}.
Additionally, in this framework all atomic interactions are modelled globally, meaning that the learning problem is solved without any inherently non-unique atom-wise, pairwise or any other many-body partitioning. Thus, the approach preserves the many-body nature of the quantum problem. 
These central properties contribute to the ability of the sGDML model to learn complex PES for molecules of intermediate size from limited amounts of reference calculations, an unachievable task for non-dedicated molecular FFs or even other ML methodologies. 
In particular, the sGDML model is able to reconstruct CCSD(T)-quality FFs from a limited amount (few hundreds) of reference molecular configurations~\cite{sgdml}.

In this chapter, we present an overview of the sGDML model from the construction of reliable datasets to the training and validation of the models to performing analysis of some relevant quantum effects captured by the model.
The structure of the chapter is the following. In section \ref{sec:DataAndSampling} we present the problem of imbalanced database and the idea behind the representative sampling.
In section \ref{sec:sGDMLmodel} we introduce the idea of physically inspired ML-FFs and present the sGDML model as well as a comparative analysis of the differences between energy and force models. 
The evaluation of the performance of the model is presented in section \ref{sec:sGDMLtests}.
Section \ref{sec:Smooth} is dedicated to the analysis of smoothing of the PES by increasing the level of theory. 
In section \ref{sec:LearnedInterac} the different type of interactions captured learned by the sGDML are highlighted. 
Finally, section \ref{sec:concl} we summarize and present the conclusions. 

\section{Data generation and Sampling of the PES}\label{sec:DataAndSampling} %
\begin{figure}[t]
\centering
\includegraphics[width=1.0\columnwidth]{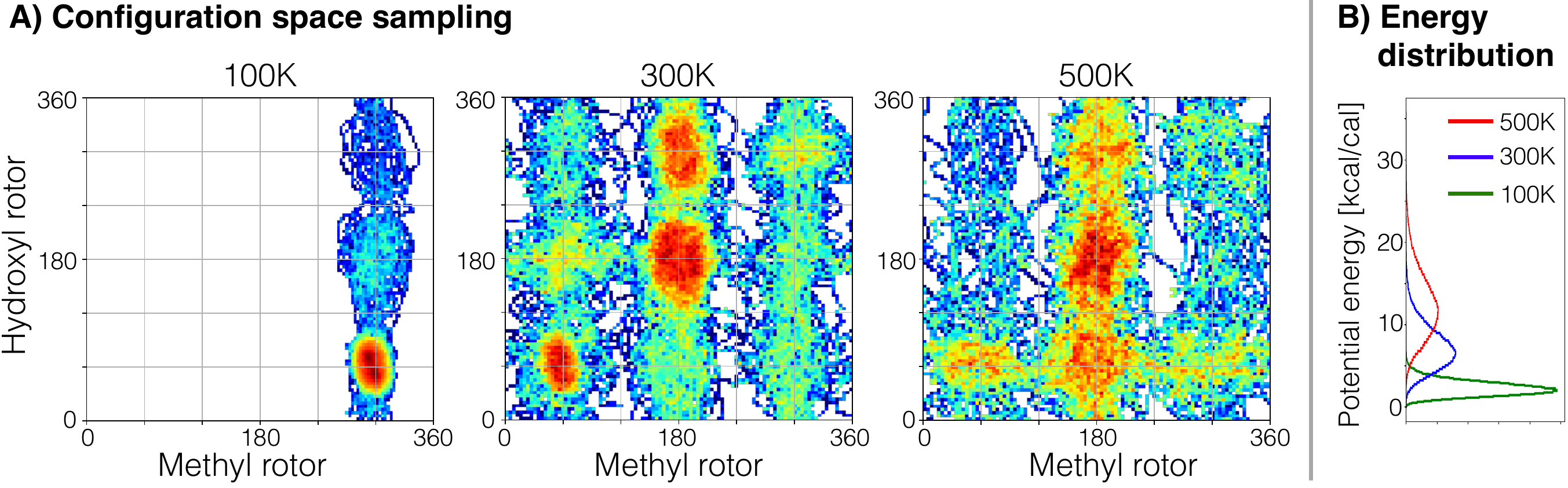}
\caption{A) Sampling of ethanol's PES at 100K, 300K and 500K using AIMD at DFT/PBE+TS level of theory. B) The potential energy profile is shown in for the different temperatures.}
\label{fig:FigEthSampl}
\end{figure}

The accurate reconstruction process of a high dimensional surface via ML methods heavily relies on the available reference data. 
In the case of PES learning, a well known approach is to construct the database by sampling the PES using molecular dynamics simulations. 
Of course the data generated with this methodology will depend on the temperature of MD simulation, therefore higher temperatures will explore higher energy regions (see Fig.~\ref{fig:FigEthSampl}).
MD-generated database will be biased to lower energy regions of the PES, where the system spend most of the time.
%Consequently, this methodology is highly advisable when the final application involves MD simulation for equilibrium or close to equilibrium properties.
Consequently, this methodology is advisable only when the final application involves MD simulation for equilibrium or close to equilibrium properties where rare events do not play a major role.
Examples of this is the study of vibrational spectra, direct study of minima population, thermodynamic properties, etc.
A general rule of thumb is to generate the database at a higher temperature compared to the intended use of the ML model trained on this data. 
For example, if we want to calculate the vibrational spectrum for ethanol at 300K, generating the database at 500K is a safe option since the subspace of configurations relevant at 300K is contained in this database (see Fig.~\ref{fig:FigEthSampl}-A). 

The main databases used in this study were created by running AIMD (DFT) simulations at a temperature of 500K using the FHI-aims package~\cite{FHIaims2009} at the generalized gradient approximation (GGA) level of theory with Perdew-Burke-Ernzerhof (PBE)~\cite{PBE1996} exchange-correlation functional and the Tkatchenko-Scheffler (TS) method~\cite{TS} to account for van der Waals interactions using the light basis set. In the literature this is known as the MD17 dataset\cite{gdml}.
%

% ==================== F I G U R E ===================
% =====================================================

\subsection{Imbalanced sampling}

\begin{figure}[t]
\centering
\includegraphics[width=1.0\columnwidth]{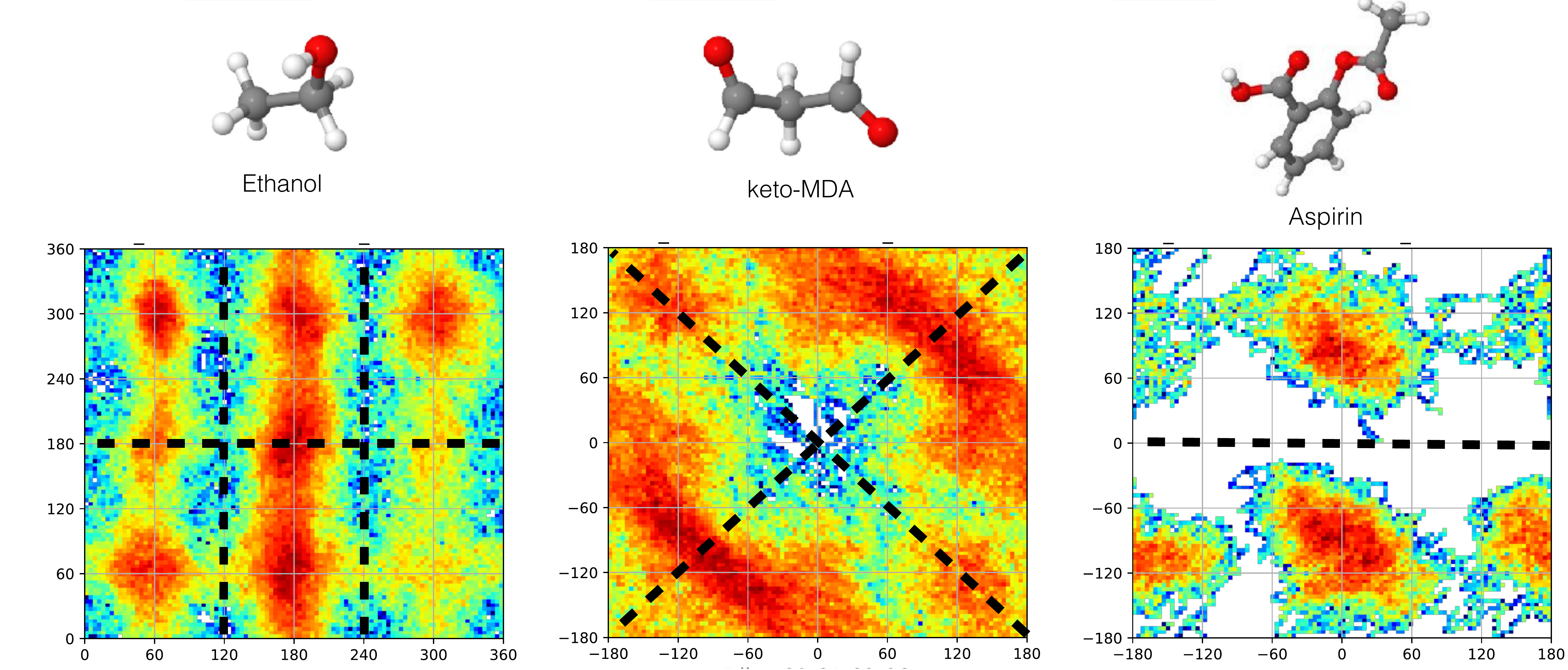}
\caption{Molecular dynamics' sampling of PESs for ethanol, keto-MDA and aspirin at 500K using DFT/PBE+TS level of theory. The black dashed lines indicate the symmetries of the molecule.}
\label{fig:FigBiasedSampl}
\end{figure}

From the ergodic hypothesis we know that the expected value of an observable $A$ can be obtained from $\langle A\rangle_{time}=N_t^{-1}\sum_{t}^{N_t}{A(\textbf{x}_t)}$ where $\textbf{x}_t$ is the step $t$ of the dynamics trajectory. This, of course, is valid only in the case in which the dynamics are long enough to visit all the possible configurations of the system under the given constraints.
%, and the employed thermostat algorithm can reproduce canonical assemble.
In practice, and in particular for AIMD this is not feasible due to its computational demands, therefore in the context of databases generation this leads to biased databases. 
Fig.~\ref{fig:FigBiasedSampl} displays the sampling of the PESs for ethanol, keto form of malodialdehyde (keto-MDA) and Aspirin at 500K using AIMD. 
It is easy to notice that even at high temperatures and more than 200 ps of simulation time, the sampling is biased and non-symmetric in the case of ethanol and Aspirin.
%
% ==================== F I G U R E ===================
% =====================================================

It is imperative to mention that when creating such databases and using them for generating ML models, many of the limitations of the database will be passed to the learned model. 
Then, the final user of ML model has to be aware of its range of applicability.
On the other hand, a robust ML framework would be able to remove some of the imperfections of the data by using prior information of the underlying nature of the data.
As an example, if training a ML-FF using the ethanol's or Aspirin's data in Fig.~\ref{fig:FigBiasedSampl} the ML model must be able to handle non-symmetric databases. Usually this is done by incorporating the indistinguishability between atoms of the same species.

\subsection{Representative sampling: From DFT to CCSD(T)}
\begin{figure}[t]
\centering
\includegraphics[width=1.0\columnwidth]{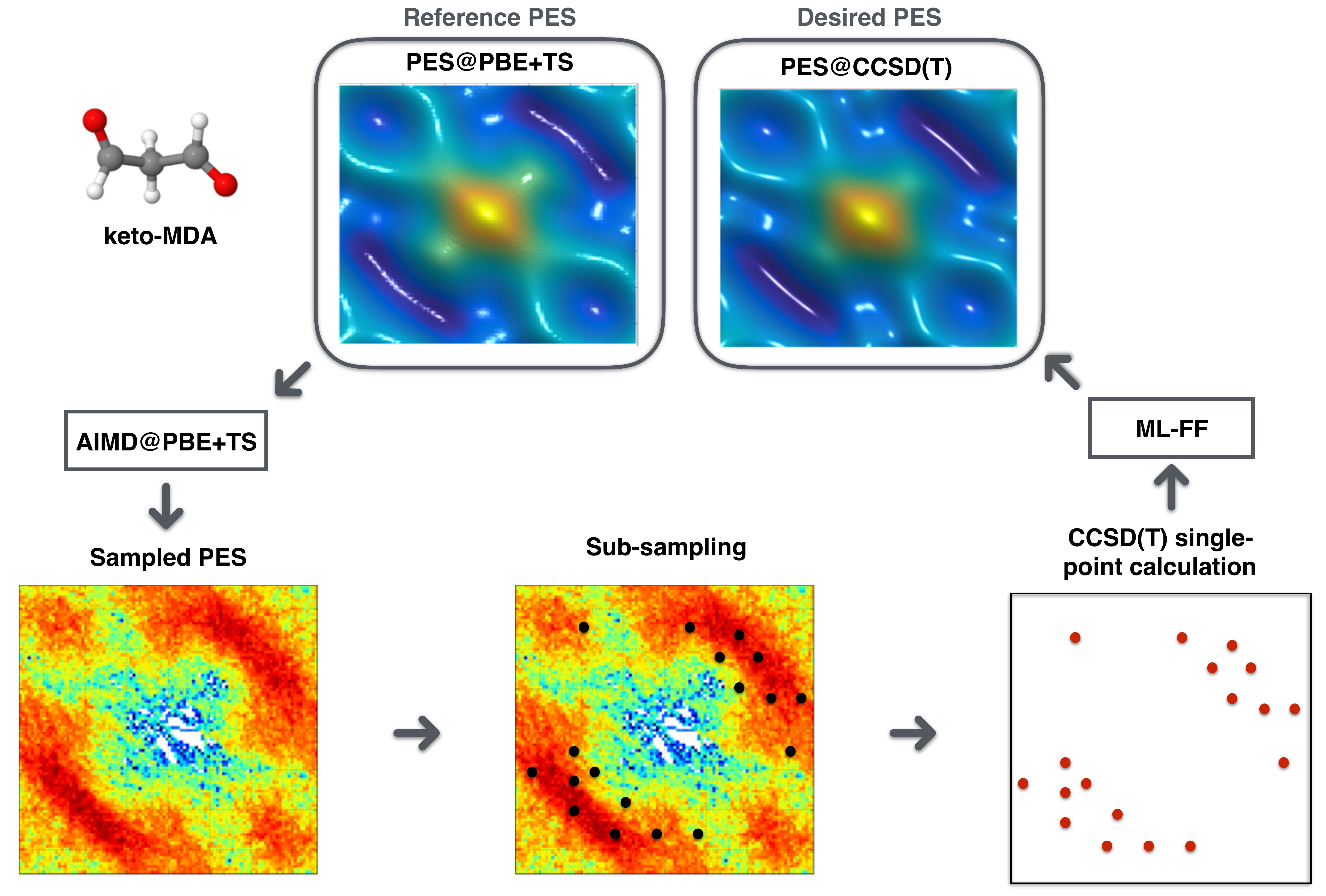}
\caption{Procedure followed to generate CCSD(T) database for the keto-MDA molecule. An AIMD simulation at 500K using DFT/PBE+TS level of theory was used as a reference sampling of the molecular PES. Afterwards the obtained trajectory is sub-sampled (black dots) to generate a subset of representative geometries, then this is used to perform single-point calculations at a higher level of theory (red dots). In this case, CCSD(T) was the desired PES and the ML-FF used was sGDML.}
\label{fig:FigSubSampl}
\end{figure}

Constructing reliable molecular databases can be very complicated even for small molecules, since efficiently exploring the molecular PES not only depends on the size of the molecule but also on many other molecular features such as intramolecular interactions and fluxional groups.
Generating $\sim$2x10$^5$ conformations from AIMD using a relatively affordable level of theory (e.g. PBE+TS with a small basis) can take from a couple of days to a couple of weeks.
Higher levels of theory (e.g., PBE0+MBD) would require weeks or months of server time. 
Finally, whenever the system under study demands the use of highly accurate methodologies such as CCSD(T), generating an extensive database becomes computationally prohibitively expensive.
To resolve this issue one can first sample the PES using a lower but representative level of theory in the AIMD simulations to generate trajectory $\{X^{\text{PBE+TS}}_t\}_{t=1}^{N_{\text{steps}}}$,  and then sub-sample this database to generate a representative set of geometries. These geometries serve as an input for higher level of theory single-point calculations, \textit{e.g.} $\{X^{\text{CCSD(T)}}_t\}_{t=1}^{N_{\text{sub-sample}}}$ (represented by red dots in Fig.~\ref{fig:FigSubSampl}), resulting in accurate and computationally affordable database.

From Fig.~\ref{fig:FigSubSampl} we see that, in this 2D projection, the reference and the desired PESs look similar, which allows to use a PES@PBE+TS sampling as a good approximation to the one that we would get by sampling PES@CCSD(T) directly~\cite{sgdml}. 
This is a crucial concept that should be carefully used since even if the test error of the ML model is good, that doesn't mean that the predictions generated by the ML model will be physically valid.
This would be the case in which the reference data comes from a PES that considerably differs from the desired PES, for example the combination of HF and CCSD(T).
Another example, when the reference data does not provide a reliable ML model, is the use of a database generated by an AIMD trajectory at 100~K for training a ML-FF, and then running MD simulations with this FF at higher temperatures. The problem is that the ML model will be generating predictions in the extrapolation regime, and therefore, there is no certainty that the results would be physically valid. 
%

% ==================== F I G U R E ===================
% =====================================================

For building sGDML models, the CCSD and CCSD(T) databases were generated by using the sub-sampling scheme (Fig.~\ref{fig:FigSubSampl}) for some of the molecules from the MD17 database. 
In the case of keto-MDA, enol-MDA and ethanol, the molecular configurations were recomputed using all-electron CCSD(T), while in the case of Aspirin all-electron CCSD were employed~\cite{psi42012,psi42017,psi42018}.

\section{Physically inspired machine learned force fields}
\label{sec:sGDMLmodel}
\sectionmark{sGDML model}
% Always give a unique label
% and use \ref{<label>} for cross-references
% and \cite{<label>} for bibliographic references
% use \sectionmark{}
% to alter or adjust the section heading in the running head
%========================== THEORY ========================== 
Machine learning offers a wide variety of different universal approximators to reconstruct any function in the limit of data availability. In practice, the amount of accessible data is restricted, specially when reconstructing complex PESs from highly accurate reference calculations such as CCSD(T).
Consequently, it is highly advantageous to mathematically constrain the space of solution of our approximator by enforcing universal physical laws, therefore naturally creating a data-efficient model capable of delivering physically meaningful predictions.
Below we summarize the desirable properties for a machine-learned force field from the physics and computational point of view:

\vskip 0.5cm
\textbf{Physical properties:}
\begin{itemize}
    \item \textit{Global model}. Building this property in the model will keep the many-body nature of the quantum interactions resulting from the solution of the Sch{\"o}dinger equation $\mathcal{H}\Psi=E\Psi$ and from the evaluation of the Hellmann-Feynman forces $-\textbf{F}=\langle \Psi^*|\partial \mathcal{H} /\partial \textbf{x}|\Psi \rangle$. In practice, this means to avoid the non-unique partitioning of the total energy $V_{BO}$ in atomic contributions. 
    \item \textit{Temporal symmetry.} This constraint demands that the ML generated Hamiltonian $\mathcal{H}= \mathcal{T}+\hat{f}_E$, with  $\mathcal{T}$ and $\hat{f}_E$ the kinetic and potential energies respectively, must be time-invariant, which means that the fundamental law of energy conservation has to be enforced in the ML model, $\mat{\hat{f}}_{\mat{F}} = - \nabla \hat{f}_E$.
    \item \textit{Indistinguishability of atoms.} In quantum mechanics, two atoms of the same specie can not be distinguished.\footnote{Even though this is a fundamental property of quantum systems, the invariance of the energy to permutations of atoms of the same species is preserved even in classical mechanics. As will be the case in all the examples discussed in this chapter.} This means that permuting two identical atoms in a molecule does not change the enegy of the system: $V_{BO}(\ldots,{\color[rgb]{0.000111,0.001760,0.998218}\vec{x}_i},\ldots,{\color[rgb]{0.986246,0.007121,0.027434}\vec{x}_j},\ldots)=V_{BO}(\ldots,{\color[rgb]{0.986246,0.007121,0.027434}\vec{x}_j},\ldots,{\color[rgb]{0.000111,0.001760,0.998218}\vec{x}_i},\ldots)$. This spatial symmetry often represents a big challenge for ML global models, but it is trivially fulfilled by models that learn energy per atom.
\end{itemize}

Each one of the above mentioned physical properties of a quantum system constitute a constraint that narrows the space of solutions of the universal ML approximator down, contributing to a more efficient and accurate reconstruction of the original data generator.

\vskip 0.5cm
\textbf{Computational requirements:}
\begin{itemize}
    \item \textit{Accuracy and data efficiency.} This is a highly desirable requirement in the reconstruction of PES from \textit{ab initio} data since the generation of each data point constitute a considerable computational cost. As an example, a CCSD(T) single-point force calculation can take several days in a single processor for a medium sized molecule. 
    \item \textit{Robust and stable predictions.} 
    To minimize the chance of artifacts in the recon- struction of the PES, the solution needs to be derived from a hypothesis space that satisfies the fundamental physical laws. Models that start from a general set of assumptions can not be expected to generalize from small data sets.
    \item \textit{Fast evaluation.} The main purpose of ML-FFs is their use in PES sampling techniques such as MD or Monte Carlo. This requires fast evaluations (few milliseconds per single-point energy/force calculations).
\end{itemize}

Whenever a ML model does not fulfill at least one of the properties or requirements mentioned above, it becomes either unreliable or inefficient for practical applications.

\subsection{Symmetrized gradient-domain machine learning}

%Gradient-domain kernel ridge regression is one of the approaches which can fulfill all the properties discussed in the previous sub-section.
Gradient-domain machine learning (GDML) is one of the approaches that fulfills all the properties discussed previously. The key idea is to use a Gaussian process (GP) to model the force field $\hat{\mathbf{f}}_{\mathbf{F}}$ as a transformation of an unknown potential energy surface $\hat{f}_{E}$ such that,
\begin{equation}
\boldsymbol{\mathbf{\hat{f}_{F}}}=-\nabla \hat{f}_{E}\sim\mathcal{G}\mathcal{P}\left[-\nabla\mu_{E}(\mathbf{x}),\nabla_{\mathbf{x}}k_{E}\left(\mathbf{x},\mathbf{x}^{\prime}\right)\nabla_{\mathbf{x}^{\prime}}^{\top}\right] \text{,}
\end{equation}
where $\mu_{E}$ and $k_{E}$ are the mean and covariance of the energy GP, respectively~\cite{LearnLinNIPS2002}. Furthermore, the model is symmetrized (sGDML) to reflect the indistinguishability of atoms, while retaining the global nature of the interactions.
With the inclusion of a descriptor $\mathbf{D:\mathcal{X}\rightarrow}\mathcal{D}$ as representation of the input, it takes the form
\begin{equation}
\hat{\mathbf{f}}_{\mathbf{F}}(\mathbf{x})=\sum_{i}^{M} \sum_{q}^{S}\mathbf{P}_q\mathbf{\alpha}_i\mathbf{J}_{\mathbf{D}(\mathbf{x})} \mathbf{k_{F}}\left(\mathbf{D}(\mathbf{x}),\mathbf{D}(\mathbf{P}_q\mathbf{x}_{i})\right)\mathbf{J}_{\mathbf{D}(\mathbf{P}_q\mathbf{x}_{i})}^{\top},
\end{equation}
where $\mathbf{J}_{\mathbf{D}(\mathbf{x})}$ is the Jacobian of the descriptor, $M$ is the number or training data points, $\mathbf{P}_q$ is the $q$th permutation in the molecular permutational group and $S$ is the size of the group. The parameters $\mathbf{\alpha}_i$ are the ones to be learned during the training procedure. 
Due to linearity, the corresponding expression for the energy predictor can be simply obtained via (analytic) integration. It is generally assumed that overly smooth priors are detrimental to data efficiency, even if the prediction target is in fact indefinitely differentiable. For that reason, (s)GDML uses a Mat\'ern kernel $k_{E}(\mathbf{x},\mathbf{x}')$ with restricted differentiability to construct the force field kernel function,
\begin{equation}
\begin{split}
\boldsymbol{k_{F}}(\boldsymbol{\mathbf{x}},\boldsymbol{\mathbf{x}}')	& =\nabla_{\mathbf{x}}k_{E}\left(\mathbf{x},\mathbf{x}^{\prime}\right)\nabla_{\mathbf{x}^{\prime}}^{\top} \\
	& =\left(5\left(\mathbf{x}-\mathbf{x}^{\prime}\right)\left(\mathbf{x}-\mathbf{x}^{\prime}\right)^{\top}-\mathbb{I}\sigma(\sigma+\sqrt{5}d)\right)
	\cdot\frac{5}{3\sigma^{4}}\exp\left(-\frac{\sqrt{5}d}{\sigma}\right)
\end{split}
\end{equation}	
Instead of using directly the molecular coordinates as representations of the system, a descriptor is used to facilitate the learning procedure. In general, it is a non-linear transformation fulfilling a set of required invariances. Here, the geometry of the molecule is represented in terms of inverse distances between all atom pairs
\begin{equation}
D_{ij}=\left\{ \begin{array}{ll}
{\left\Vert \mathbf{r}_{i}-\mathbf{r}_{j}\right\Vert ^{-1}} & {\text{ for }i>j}\\
{0} & {\text{ for }i\leq j}
\end{array}\right\},
\end{equation}
making the model invariant to roto-translations.

A full symmetrization of the model requires summing over all possible permutations of its inputs. To avoid the combinatorial challenge associated with summing over large symmetry groups, we restrict ourselves to the much smaller subset of physically plausible rigid space group and fluxional symmetries, $\{\mathbf{P}_q\}_{q=1}^{S}$.

Extracting those symmetries usually requires chemical and physical intuition about the system under study, e.g. rotational barriers, which is impractical in a ML setting. To automate that step, we employ a multi-partite matching scheme that identifies and recovers the permutational transformations undergone by the system within the training dataset.

This is achieved by finding the permutation operation $\tau$ that minimizes the cost function,
\begin{equation}
\operatorname*{arg\,min}_{\tau} \mathcal{L}(\tau) = \|\mat{P}(\tau)\mat{A}_G\mat{P}(\tau)\tran - \mat{A}_H\|^2,
\label{eq:matching_objective}
\end{equation}
between adjacency matrices $(\mat{A})_{ij} = \|\vec{r}_i - \vec{r}_j\|$ of all molecular graph pairs $G$ and $H$ in different energy states. A particular challenge is to find matchings that are consistent across the whole training set.

The set of permutations $\{\mathbf{P}_q\}_{q=1}^{S}$ obtained by this method, also known as the \textit{Higgins group}, omits unfeasible transformations that do not contribute any valuable information to the inference task and thus help in reducing the computational effort required to evaluate the model.

\begin{figure}[h]
\centering
\includegraphics[width=0.8\columnwidth]{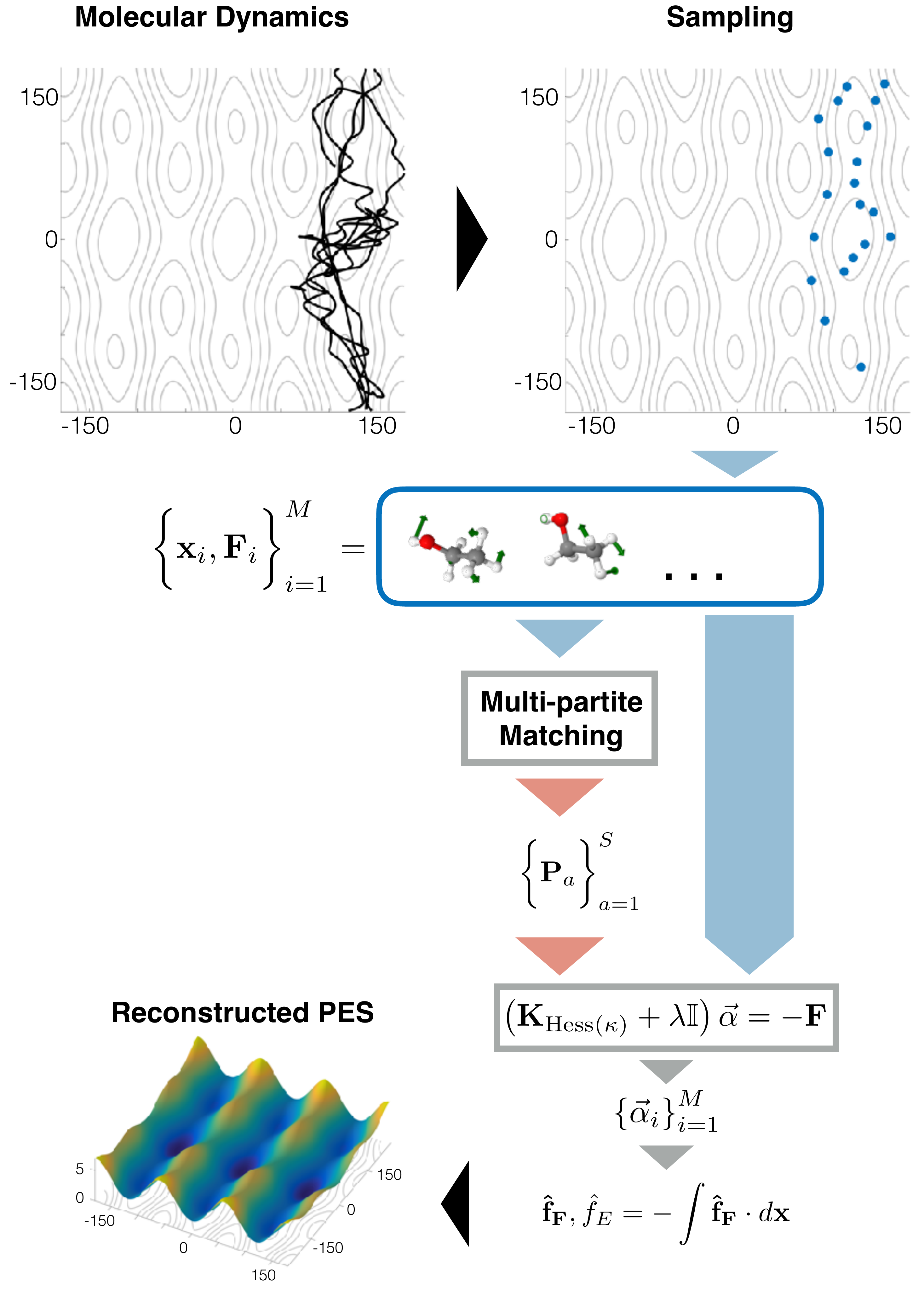}
\caption{Construction of the sGDML model. (1) The data used for training, \{\textbf{x}$_i$, \textbf{F}$_i$\}$_{i=1}^M$, is created by subsampling molecular dynamics trajectories (blue dots). The forces are represented by green arrows on top of each atom. (2) The set of molecular permutation symmetries, \{\textbf{P}$_a$\}$_{a=1}^S$, are extracted from the training set by the multi-partite matching approach. This effectively enhances the size of the training set by a factor $S$ and symmetrizes the PES. (3) The force field is trained by solving the linear system for \{$\alpha_j$\}. The reconstructed PES is obtained by analytical integration of the force predictor.}
\label{fig:FigsGDMLtrain}
\end{figure}

A sketch of the general training procedure is shown in Fig.~\ref{fig:FigsGDMLtrain}, from sampling a molecular dynamics trajectory and extracting the Higgins group to solving the normal equation and generating the embedded PES in the data.

%
% ==================== F I G U R E ===================
% =====================================================

%In a series of recent publications
In Refs.\cite{gdml,sgdml,sGDMLjcp,sGDMLsoftware2019} it was demonstrated that the sGDML framework is highly data efficient being able to achieve state-of-the-art predictions even when trained on only a few hundred reference data points.
As examples, it is possible to reconstruct molecular PESs with a mean absolute error of less that 0.06 kcal~$\text{mol}^{-1}$ for small molecules (e.g. with up to 15 atoms) and 0.16 kcal~$\text{mol}^{-1}$ for more complex molecules (e.g. aspirin, paracetamol, and azobenzene)~\cite{sgdml}. 
Such accuracy is achieved while following physical requirements and therefore resulting in robust learning models which are capable of decoding complex subtleties hidden in the reference data.
\subsection{Force vs. energy ML models}

As stated in the previous section, the sGDML framework is constructed for being trained in the gradient domain of the energy function. This approach contrasts with conventional ML methodologies based on direct energy function learning (using energies and forces for training) in which the forces are computed via analytic differentiation~\cite{Alder1959,Rahman1964,Verlet1967,Rahman1971,EAM1984,Tersoff1988,PolarFF2006,TIP4P1983,TIP5P2000,AMBER1981,CHARMM1983,MMFF94,GROMOS2005,Behler2007,Behler2007b,Behler2011,Behler2011a,Behler2012,Behler2016,Gastegger2017,SchNetNIPS2017,SchNet2018,Bartok2013,Bartok2015_GAP,Ramprasad2015,Rupp2015,Glielmo2017}, as represented in the next diagram:

\begin{equation*}
  \begin{CD}
    &\text{Trained} & & \text{Derived} \\
    \text{sGDML}:&\mat{\hat{f}_F} @>>> \hat{f}_{E \gets \mat{F}} = -\int \mat{\hat{f}_F} \cdot d\mat{x}+C \\
    \text{E-ML}: &\hat{f}_E  @>>> \mat{\hat{f}}_{\mat{F} \gets E} = - \nabla \hat{f}_E
  \end{CD}
\end{equation*}

\noindent where E-ML refers to energy machine learned models.

Any ML model has an associated learning uncertainty~\cite{sGDMLuncertainty}. This uncertainty is also present during the evaluation of the model. 
Given the nature of the operations in obtaining the derived quantities in the previous diagram, we can see that there is an advantage in learning the force field directly over the energy models.
Lets consider the ensembles of models \{$\mat{\hat{f}_F}$\} and \{$\hat{f}_E$\} with mean $\langle\mat{\hat{f}_F}\rangle$ and $\langle\hat{f}_E\rangle$ and uncertainties $\gamma_F$ and $\gamma_E$, respectively. 
It can be shown that, in the case of the sGDML model the uncertainty that propagates from the ensemble to the ensemble of energies $-\int \mat{\hat{f}_F} \cdot d\mat{x}$ is given by $\sim\gamma_F\Delta x$ where $\Delta x$ is a small number in the length scale.
In the case of the uncertainty in the derived forces from E-ML, $-\nabla \hat{f}_E$ is given by $\sim\gamma_E / \Delta x$. 
From this simple analysis we conclude that: \textit{the error attached to energies from the sGDML model will be attenuated while errors in predicted forces from E-ML models will be amplified}~\cite{sGDMLjcp,ChmielaThesis}.
Another intuitive proof of this effect was reported from signal processing theory  point of view in the GDML original article~\cite{gdml}. 
This fundamental result highlights the irrefutable advantage of gradient domain learning over energy based learning, which evince the robustness and stability of such ML framework.

\section{Gradient domain learning and its performance} \label{sec:sGDMLtests}

In this section we analyze the performance of the sGDML framework in reconstructing molecular force fields and their underlying potential energy surfaces. 
First, from the point of view of cross validation which judge its ability to predict unseen data, and second, perhaps a more physically relevant validation, a direct comparison with the reference method (e.g. DFT) of statistical properties computed from molecular dynamics simulation. 
%This study was done for a series of molecules showed in Fig.~\ref{fig:FigContent}.

\subsection{Static validation}
Table~\ref{tab:results_all_sym} shows the sGDML prediction results for six molecule datasets trained on 1000 geometries, sampled uniformly according to the MD@DFT trajectory energy distribution (see Fig.~\ref{fig:FigsGDMLtrain}).
It is easy to notice that for all the considered molecules the mean absolute error (MAE) in the energies is below 0.2 kcal mol$^{-1}$, and even lower than 0.1 kcal mol$^{-1}$ for small molecules. Remarkable achievement considering that the model was trained using only 1000 training data points.
% 
% ________________________________________
% _____________ TABLE ____________________|
%
\begin{table*}[!ht]
\centering
\caption{Prediction accuracy for total energies and forces of the sGDML$@$DFT. The mean absolute errors (MAE) and root mean squared error (RMSE) for the energy and forces are in kcal $\text{mol}^{-1}$ and kcal $\text{mol}^{-1} \text{\AA}^{-1}$, respectively. These results were originally published in Refs.\cite{sgdml} and \cite{sGDMLjcp}.}
\begin{threeparttable}
\label{tab:results_all_sym}
\setlength\extrarowheight{1pt}
%\begin{ruledtabular}
\begin{tabular}{l|l|rr|rr|rrrr}
\multicolumn{2}{l}{Dataset} & \multicolumn{2}{c}{Energies} & \multicolumn{4}{r}{Forces}\\[2pt]
\cline{1-2} \cline{3-4} \cline{5-10}
\\ [-2.3ex]
&&&&&&\multicolumn{2}{c}{Magnitude} & \multicolumn{2}{c}{Angle} \\[2pt]
\cline{7-8} \cline{9-10}
 \\ [-2.3ex]
Molecule & \# ref. & MAE & RMSE & MAE & RMSE & MAE & RMSE & MAE & RMSE \\[1pt]
\hline \\ [-2.3ex]
keto--MDA & % 1000, e-15, 58
\ensuremath{1000} &
\ensuremath{0.10} &
\ensuremath{0.13} &
\ensuremath{0.41} &
\ensuremath{0.62} &
\ensuremath{0.39} &
\ensuremath{0.56} &
\ensuremath{0.0055} &
\ensuremath{0.0087}\\
Ethanol & % 1000, e-15, 76
\ensuremath{1000} &
\ensuremath{0.07} &
\ensuremath{0.09} &
\ensuremath{0.33} &
\ensuremath{0.49} &
\ensuremath{0.46} &
\ensuremath{0.63} &
\ensuremath{0.0051} &
\ensuremath{0.0083}\\
Salicylic acid & % 1000, e-15, 38
\ensuremath{1000} &
\ensuremath{0.12} &
\ensuremath{0.15} &
\ensuremath{0.28} &
\ensuremath{0.44} &
\ensuremath{0.32} &
\ensuremath{0.45} &
\ensuremath{0.0038} &
\ensuremath{0.0064}\\
enol--MDA & % 1000, e-15, 58
\ensuremath{1000} &
\ensuremath{0.07} &
\ensuremath{0.09} &
\ensuremath{0.13} &
\ensuremath{0.22} &
\ensuremath{-} &
\ensuremath{-} &
\ensuremath{-} &
\ensuremath{-}\\
Paracetamol & % 1000, e-15, 42
\ensuremath{1000} &
\ensuremath{0.15} &
\ensuremath{0.20} &
\ensuremath{0.49} &
\ensuremath{0.70} &
\ensuremath{0.60} &
\ensuremath{0.84} &
\ensuremath{0.0073} &
\ensuremath{0.0118}\\
Aspirin & % 1000, e-15, 68
\ensuremath{1000} &
\ensuremath{0.19} &
\ensuremath{0.25} &
\ensuremath{0.68} &
\ensuremath{0.96} &
\ensuremath{0.52} &
\ensuremath{0.68} &
\ensuremath{0.0094} &
\ensuremath{0.0139}\\
\end{tabular}
%\end{ruledtabular}
\end{threeparttable}
\end{table*}
%
% ________________________________________
% ________________________________________|
%
This contrasts with pure energy-based models (e.g. other kernel models~\cite{gdml} or neural networks~\cite{Cormorant2019}) which require up to two orders of magnitude more samples to achieve a similar accuracy.
As shown in the original GDML article~\cite{gdml}, the superior performance of gradient based learning cannot be simply attributed to the greater information content of force samples (one energy value per $3N$ force components per sample). 
Lets consider a direct comparison of two kernel models, energy and gradient based, for energy learning with the same number of degrees of freedom (non symmetrized versions for simplicity),
\begin{equation}
-\hat{f}_{E\gets\mat{F}}(\vec{x})  = \sum^M_{i} \{
{\color[rgb]{0.000111,0.001760,0.998218}\vec{\alpha}_{i} \cdot \nabla}\}\kappa(\vec{x},\vec{x}_i)\,.
%\label{eq:force_model}
\end{equation}
\begin{equation}
-\hat{f}_E(\vec{x}) = \sum^{3N\times M}_{j}{\color[rgb]{0.000111,0.001760,0.998218}\beta_{j}}\kappa(\vec{x},\vec{x}_j)\,.
%\label{eq:force_model}
\end{equation}
Then, each model has $3N\times M$ parameters with the difference that, in the energy model the $\{\beta_{j}\}^{3N\times M}_{j=1}$ parameters are correlation only by the learning procedure, while in the force model exist the additional correlation imposed in the triads $\{\alpha^x_{i},\alpha^y_{i},\alpha^z_{i}\}^{N\times M}_{i=1}$ by the gradient operator.
Hence, this extra correlation between the parameters imposed by learning in the gradient domain reduces our space of solutions and therefore the model becomes more data efficient. 
Such fundamental characteristic positions the sGDML modes in a privileged place for learning force fields from highly accurate quantum chemical methodologies (e.g. CCSD(T)) in which data is very scarce, where even generating 100 data points is a monumental computational task. 
In the next section we will analyze this topic, but for now lets validate the sGDML models by direct comparison with MD simulations generated with the reference method.

\subsection{Dynamic validation}

In the previous section, we saw that the prediction errors in sGDML learned models are very low. Nevertheless, a natural question to ask is if the molecular dynamics simulations using the learned models (i.e. MD@sGDML\{DFT\}) can actually replicate the statistical properties of the physical system as computed running MD simulations using the \textit{ab-initio} reference theory(i.e. MD@DFT).
To address this issue, in this section we present MD simulations with sGDML and DFT forces for benzene, uracil, and aspirin molecules. All the simulations have been done within precisely the same conditions (temperature, integrator, integration step, software, etc.) using the i-PI molecular dynamics package~\cite{iPI2014}.

% ==================== F I G U R E ===================
\begin{figure}[t]
\centering
\includegraphics[width=1.0\columnwidth]{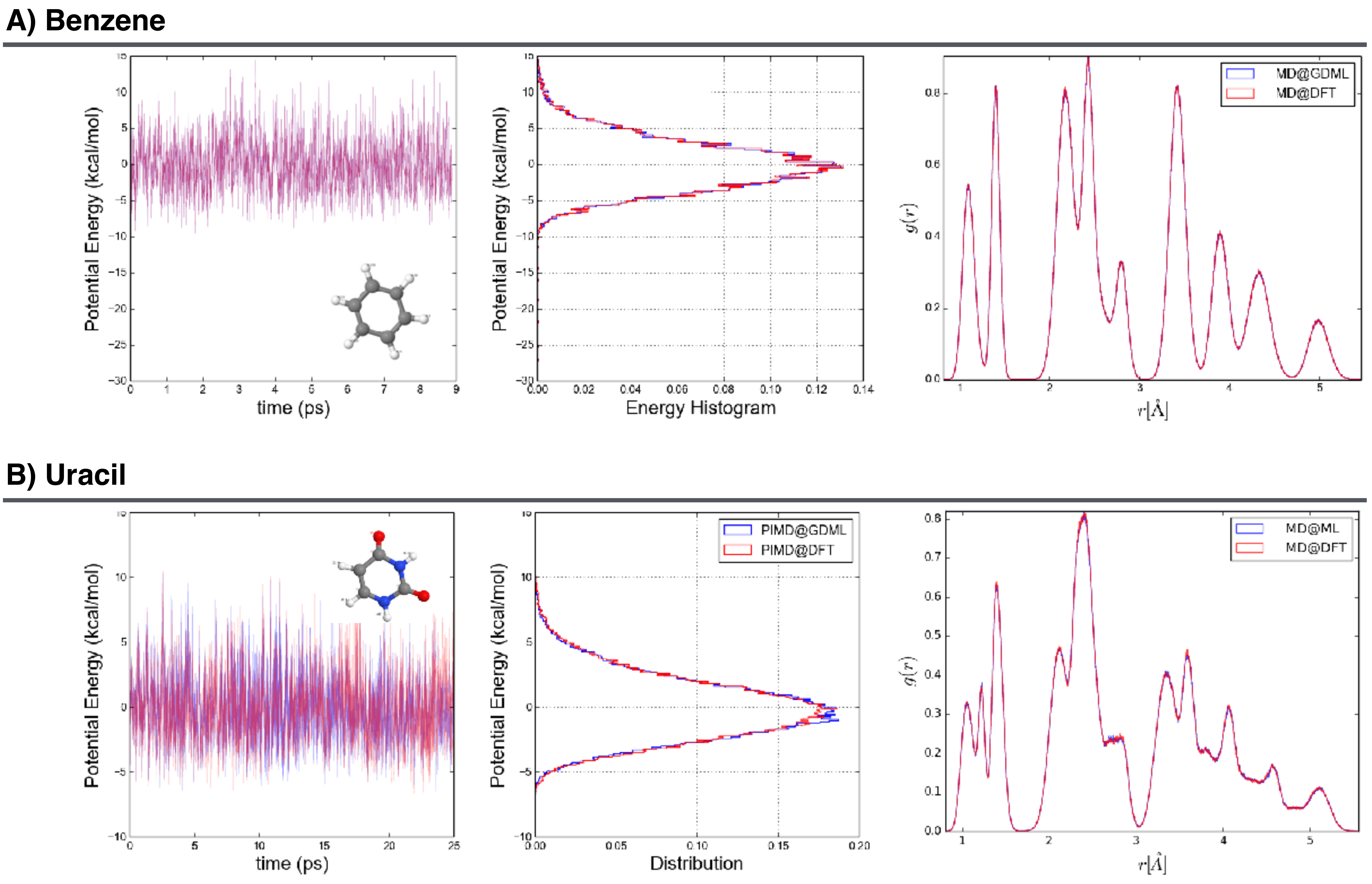}
\caption{Accuracy of potential energies (minus the mean value) sampling for sGDML$@$DFT (using PBE+TS functional) and sGDML$@$CCSD(T) models on various molecular dynamics datasets. Energy errors are in kcal $\text{mol}^{-1}$. These results, with the exception of enol-MDA, were originally published in Ref.~\cite{sgdml}. All the models were trained using atomic forces for 1000 molecular conformations.}
\label{fig:FigsGDMLvsDFT_bznUra}
\end{figure}
% =====================================================

%
\subsubsection{Benzene and uracil}

In the case of benzene, we have performed MD simulations at 300~K using the same initial conditions for both MD@DFT and MD@sGDML\{DFT\}. Fig.~\ref{fig:FigsGDMLvsDFT_bznUra}-A shows the evolution of the potential energy in time and we can see a very good agreement. 
From this we can deduce that, at least in the first 10 ps of the trajectory, a MAE of 0.1 kcal $\text{mol}^{-1}$/0.06 kcal $\text{mol}^{-1}$ \AA$^{-1}$ in energies/forces for benzene's sGDML model don't generate significant deviations from the reference MD@DFT trajectory.
In the case of uracil we repeated the same experiment but this time we started the simulations from different initial conditions and ran the simulations for 25 ps to collect more statistics. 
Fig.~\ref{fig:FigsGDMLvsDFT_bznUra}-B displays the evolution of the two potential energies, MD@DFT in red and MD@sGDML\{DFT\} in blue. It can be seen that both methods generate the same potential energy sampling (Fig.~\ref{fig:FigsGDMLvsDFT_bznUra}-B-middle) and the same interatomic distance distribution (Fig.~\ref{fig:FigsGDMLvsDFT_bznUra}-B-right). Therefore, the MAE of 0.11 kcal $\text{mol}^{-1}$/ 0.24 kcal $\text{mol}^{-1}$ \AA$^{-1}$ in energies/forces for uracil's model does not generate significant deviations from the exact reference data up to 25 ps of trajectory.

\subsubsection{Aspirin}
\begin{figure}[t]
\centering
\includegraphics[width=1.0\columnwidth]{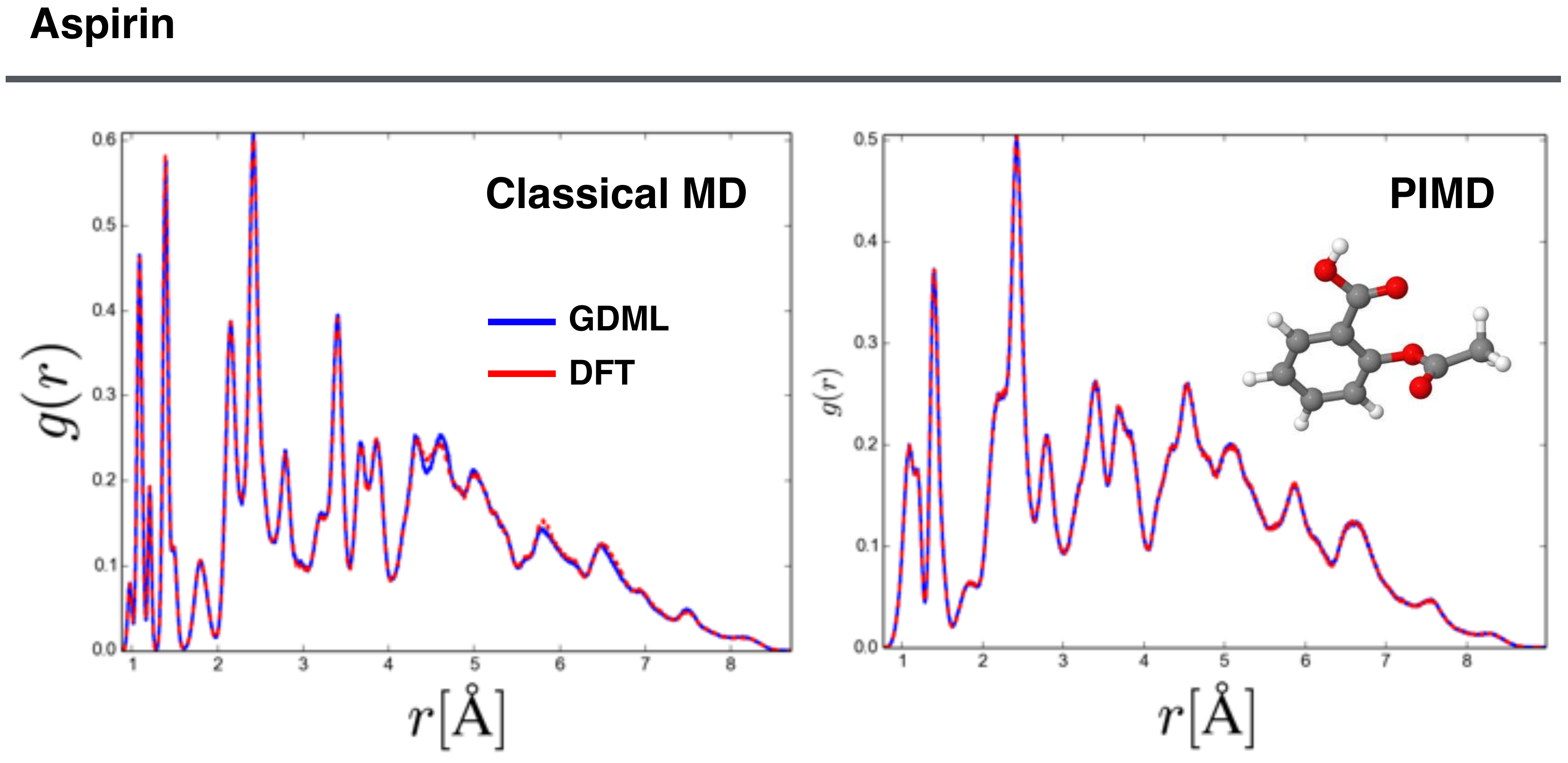}
\caption{Comparison of the interatomic distance distributions obtained from GDML (blue line) and DFT (dashed red line) with classical MD (left), and PIMD (right).}
\label{fig:FigsGDMLvsDFT_Asp}
\end{figure}

% Aspirin 
A more interesting case is aspirin, which is a much more complex molecule. In this case by running MD@GDML at 300~K, overall we observe a quantitative agreement in interatomic distance distribution between MD@DFT and MD@GDML simulations (Fig.~\ref{fig:FigsGDMLvsDFT_Asp}-left). 
The small differences can be observed only in the distance range between 4.3 and 4.7 \AA. This region mainly belongs to the distances between the two main functional groups in aspirin. 
Slightly higher energy barriers in the GDML model affect the collective motion of these groups, which results in a small difference in the interatomic distance distributions.
These differences in the interatomic distance distributions vanish once the quantum nature of the nuclei is introduced via path integral molecular dynamics (PIMD) simulations (Fig.~\ref{fig:FigsGDMLvsDFT_Asp}-right)~\cite{gdml}. 
Consequently, by running more realistic simulations we overcome the small imperfections in the reconstruction of the PES allowing to generate more accurate results. 

% ==================== F I G U R E ===================
% =====================================================

By performing \textit{static} and \textit{dynamic} validations in sGDML learned models we have demonstrated the robustness and data efficiency (the models were trained only on 1000 data points) of the framework.
In the next section, we briefly analyze interesting synergistic behavior between the data efficiency of the sGDML and using more accurate reference calculations. 

\section{Smoothness hypothesis in quantum chemistry} \label{sec:Smooth}

Within the Born-Oppenheimer approximation, the potential energy surface $V_{BO}(\mathbf{x})$ is the energy eigenvalue of the Schr\"odinger equation $\mathcal{H}\Psi=V_{BO}\Psi$, which parametrically depends on a given set of nuclear coordinates $\mathbf{x}$, and the level of theory used to approximate its solution will of course define its accuracy. 
A very basic approximation is given by the Hartree-Fock theory (HF) in which the correlation between electrons of the same spin is treated as a mean field rather than as an instantaneous interaction and the correlation between electrons of opposite spins is omitted. 
To incorporate the missing \textit{electron correlation}, other post-HF approximations were built on top of HF solutions, such as M{\o}ller--Plesset perturbation theory (e.g. MP2, MP3, and MP4), Coupled cluster (e.g. CCSD, CCSD(T), and CCSDT) and Configuration interaction (e.g. CISD and Full CI), etc.
Unfortunately, moving to more accurate approximations is associated with a steep increase in the needed computational resources, making unfeasible to perform calculations for example using Full CI for molecules such as ethanol. 
In the case of density functional theory, which is less computationally demanding, it is not clear how to hierarchically increase electron correlation by going from one exchange-correlation functional to another one. Therefore, we focus only on post-HF methods.

\begin{figure}[t]
\centering
\includegraphics[width=1.0\columnwidth]{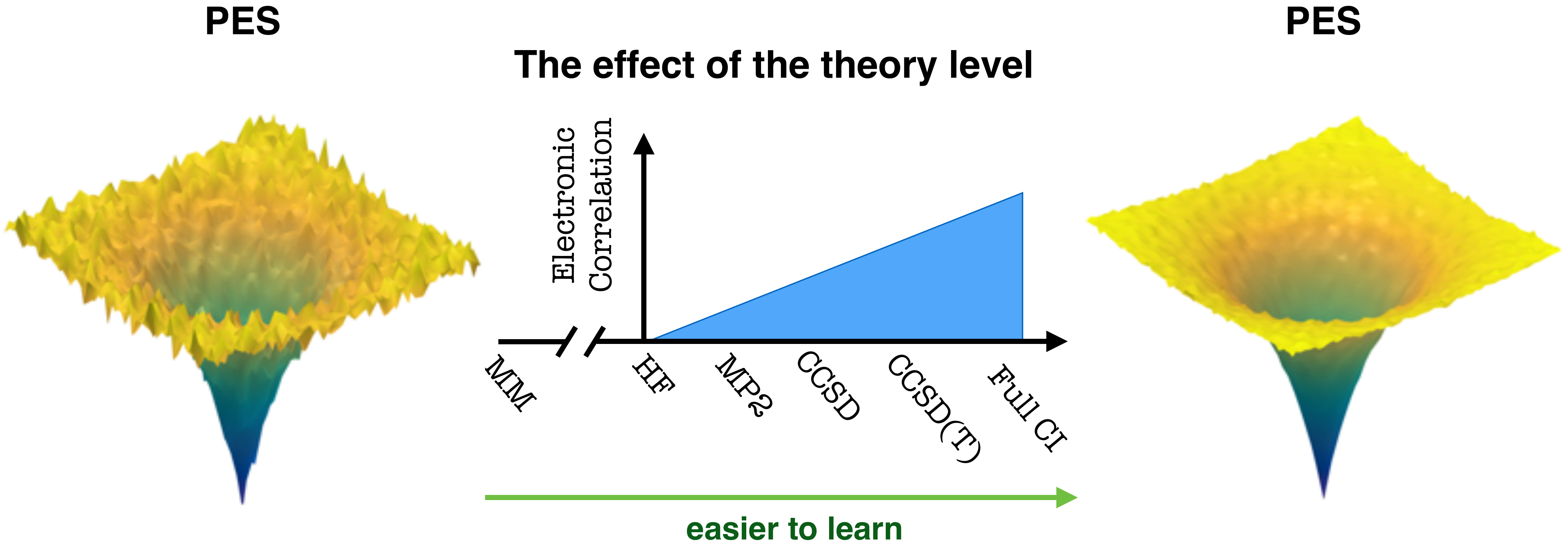}
\caption{Pictorial representation of the smoothness hypothesis in quantum chemistry.}
\label{fig:FigSmoothHyp}
\end{figure}

The smoothness hypothesis states that \textit{systematically increasing the amount of electron correlation will systematically smoothen the ground state potential energy surface} (see Fig.~\ref{fig:FigSmoothHyp}).

\begin{figure}[t]
\centering
\includegraphics[width=1.0\columnwidth]{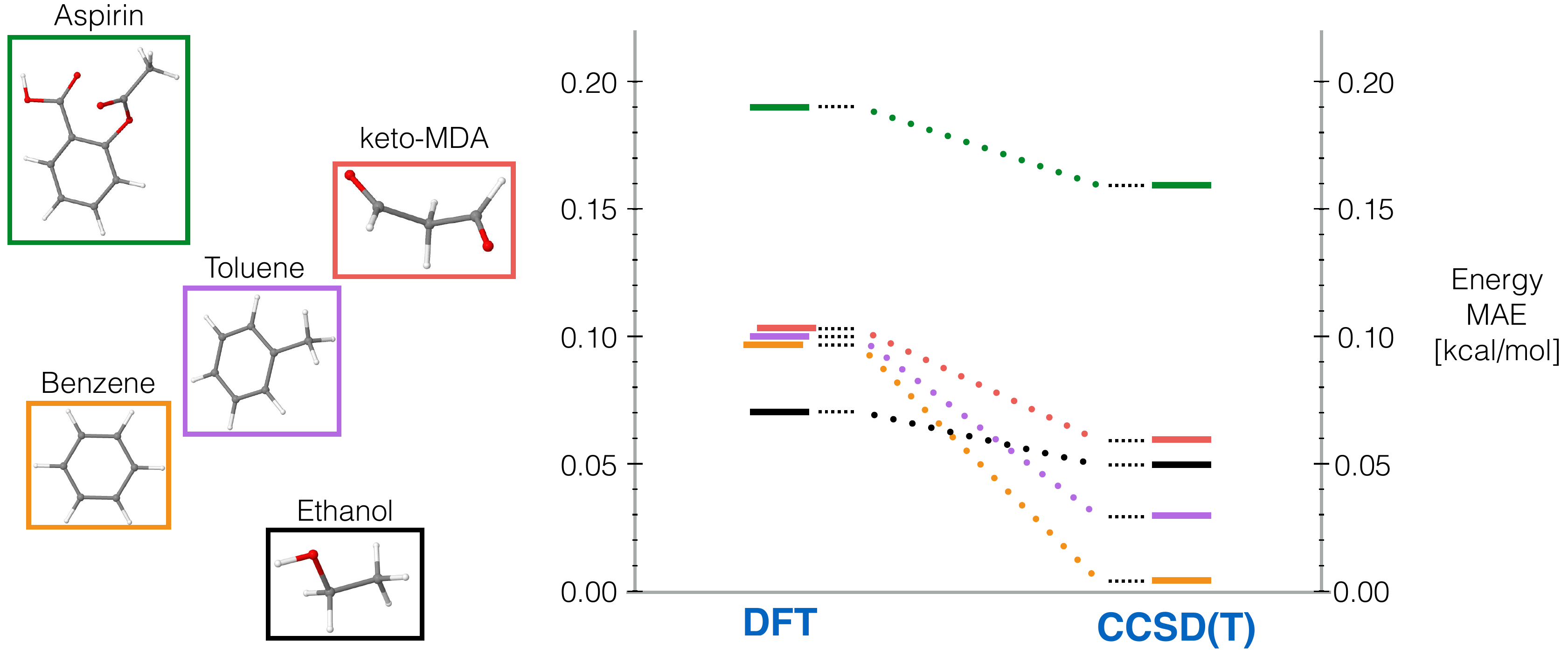}
\caption{Accuracy of total energies for sGDML$@$DFT (using PBE+TS functional) and sGDML$@$CCSD(T) models on various molecular dynamics datasets. Energy errors are in kcal $\text{mol}^{-1}$. These results, with the exception of enol-MDA, were originally published in Ref.~\cite{sgdml}. All the models were trained using atomic forces for 1000 molecular conformations.}
\label{fig:FigDFT2CC}
\end{figure}

As stated in the previous section, the sGDML framework is characterized for delivering state-of-the-art accuracies while using only a handful of training data points.
This allows to construct compact sGDML models that faithfully reconstruct molecular force fields even from computationally costly \textit{ab initio} methods such as the gold standard in quantum chemistry all-electron coupled cluster with single, double, and perturbative triple excitations (CCSD(T)).
Now, by following the procedure described in Fig.~\ref{fig:FigSubSampl} we trained a set of molecules using CCSD(T) reference data, giving very interesting results as displayed in Fig.~\ref{fig:FigDFT2CC}. For all the molecules in this study, \textit{the prediction energy error of the sGDML models dropped just by increasing the level of theory of the training data.}

Furthermore, in the case of benzene the MAE drastically reduces to only few \textit{cal mol$^{-1}$}!

From the signal reconstruction point of view, the smoother or the lower the complexity of the signal the easier to reconstruct. 
Meaning that less complex functions from the space of solutions can be used to capture the intrinsic features encoded in the reference data. 
Hence, given that increasing the electron correlation (going from DFT to CCSD(T)) makes the problem easier to learn (see Fig.~\ref{fig:FigDFT2CC}) and because of the above given argument, we can say that for the studied molecules our results support the smoothness hypothesis (Fig.~\ref{fig:FigSmoothHyp})~\cite{sGDMLjcp}.
An explanation why some molecules profit more than others by increasing the level of theory is not clear and needs further research.

% ==================== F I G U R E ===================
% =====================================================

% ==================== F I G U R E ===================
% =====================================================

%%%%%%%%%%%%%%%%%%%%%%%%%%%%%%%%%%%%%%%%%%%%%%%%%%%%%%%%%%%%%%
%%%%%%% ========================== MOLECULAR    PES ========================== %%%%%%%%
%%%%%%%%%%%%%%%%%%%%%%%%%%%%%%%%%%%%%%%%%%%%%%%%%%%%%%%%%%%%%%
\section{Learning molecular PES: What type of interactions can be captured?} \label{sec:LearnedInterac}

In this section, we exemplify the insights obtained with sGDML model for ubiquitous and challenging features of general interest in chemical physics: 
intramolecular hydrogen bonds, electron lone pairs, electrostatic interactions, proton transfer effect, and other electronic effects (e.g. bonding--antibonding orbital interaction and change in the bond nature).

The PES, $V_{BO}$, contains all the information necessary to describe the dynamics of a molecular system.
Its intricate functional form results from the interplay between different quantum interactions, characteristic that should be preserved during the learning process.
Consequently, it is not known how to expand the $V_{BO}$ in different energetic contributions (e.g. hydrogen bonding, electrostatics, dispersion interactions or other electronic effects) to make it more interpretable.
Nevertheless, by accurately learning the $V_{BO}$ at a high level of theory using the sGDML framework, we can perform careful analysis on the learned models and its results from applications (e.g. MD simulations) to decode many of the complex features contained in the quantum-chemical data.

In practice, these features or intramolecular interactions (e.g. van der Waals interactions, energy barriers or H-bond interactions) are subtle variations in the energy surface of less than 0.1 kcal~$\text{mol}^{-1}$, one order of magnitude lower than so-called chemical accuracy.
An particular example is the ethanol molecule. The relative stability of its \textit{trans} and \textit{gauche}$^{(l,r)}$ conformers is within 0.1 kcal~$\text{mol}^{-1}$. 
Furthermore, the energetic barriers \textit{trans} $\rightleftharpoons$ \textit{gauche}$^{(l,r)}$ and \textit{gauche}$^{(l)}$ $\rightleftharpoons$ \textit{gauche}$^{(r)}$ differ only by $\sim$0.1 kcal~$\text{mol}^{-1}$ too. 
%This means that the dynamics of the system would be seriously affected if those stringent accuracies are not reached~\cite{sgdml}. 
Any machine learning model with an expected error above those stringent accuracies risk misrepresenting the molecular system or even inverting this subtle energy difference, which will lead to incorrect configuration probabilities and hence quantitatively wrong dynamical properties.
The robust sGDML framework has been shown to satisfy such stringent demands, obtaing MAEs of 0.1--0.2 kcal~$\text{mol}^{-1}$ for molecules with up to 15 atoms~\cite{sgdml}.
\footnote{Even thought the MAE is in the same order as the required accuracy, we have to mention that this error is computed in the whole data-set. This means that the error in the highly sampled regions (e.g. local minima) will be lower than the reported MAE.}
Moreover, as shown in Fig.~\ref{fig:FigDFT2CC}, the prediction error can be even lower by training on coupled-cluster reference data. 
With the certainty that we are working with very accurate ML models, we can confidently analyze and interpret their results.

%%%%%%%%%%%%%%%%%%%%%%%%%%%%%%%%%%%%%%%%%%%%%%%%%%%%%%%%%%%%%%
\subsection{Electrostatic interactions and electron lone pairs} \label{QQint}

First, we focus our attention on electrostatic interactions, in particular lone-pair--atom interaction. 
The concept of electron lone pairs plays a central role in chemistry, these are ubiquitous atomic features responsible for a wide variety of phenomena.
A simple way to define lone pairs is as atomic valence electrons that are not shared with any other atom in a molecule i.e. they are not involved in bond formation. They are often present as lone pairs of nitrogen and oxygen atoms in a molecule.

\subsubsection{Electron lone pairs in ethanol}

A very illustrative case used along this chapter is ethanol molecule:
\textit{i}) it has two rotors -- hydroxyl and methyl groups -- as main degrees of freedom making very easy to visualize its PES, 
\textit{ii}) due to its complex electronic structure it requires at least CCSD(T) to correctly describe its PES, 
\textit{iii}) despite its simple appearance it is not trivial to reconstruct its force field, and 
\textit{iv}) it presents a rich variety of intramolecular interactions such as the strong effects of electron lone pairs on its dynamics. 
% ==================== F I G U R E ===================
\begin{figure}[b]
\centering
\includegraphics[width=1.0\columnwidth]{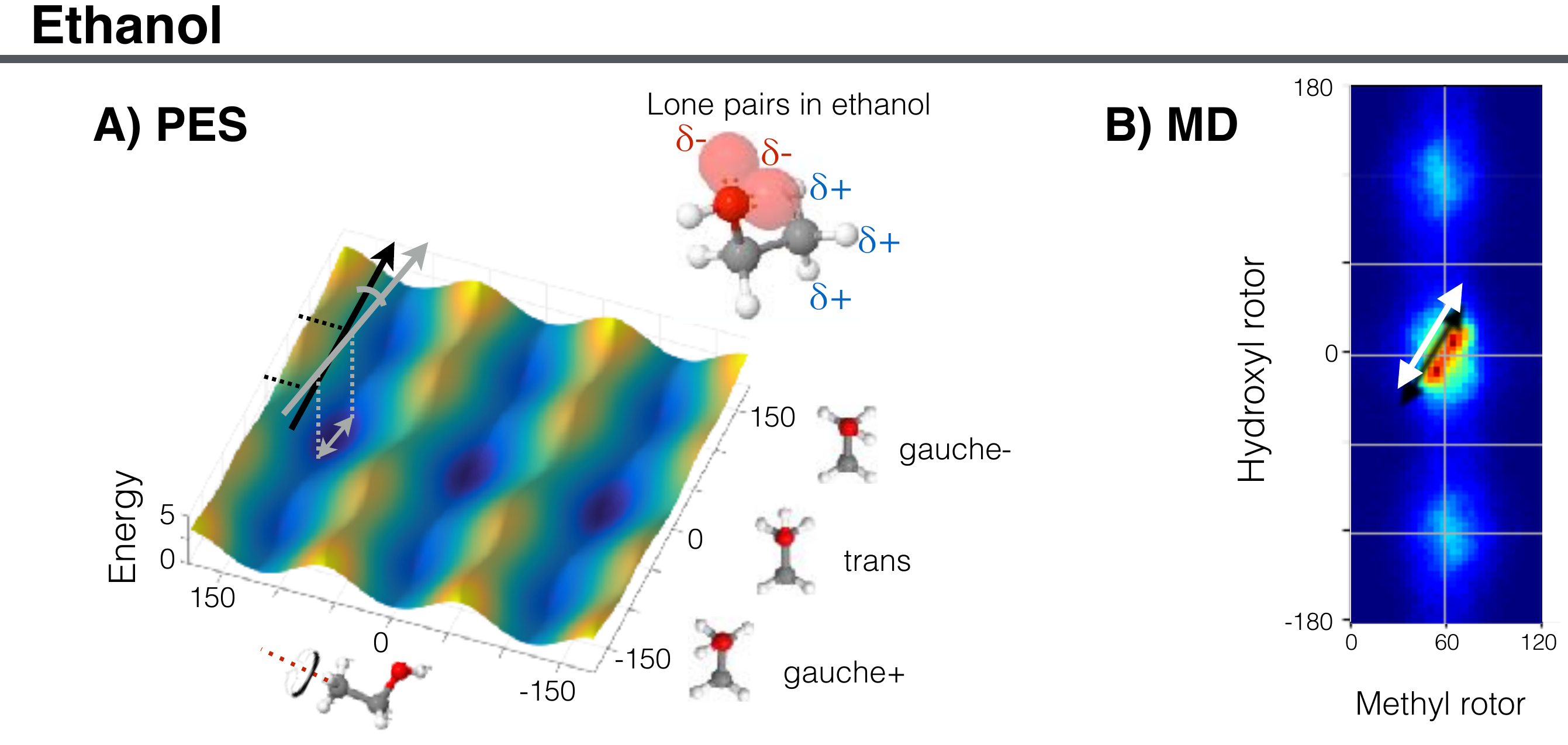}
\caption{A) Ethanol's PES (sGDML@CCSD(T) model). The molecule shows the effect of oxygen's lone pair and the partial positive charges in methyl's hydrogen atoms and their coupling is represented by a gray arrow in the PES of ethanol. B) PES sampling generated by MD@sGDML simulations at 300K using the NVE ensemble.}
\label{fig:FigEthanol}
\end{figure}
% =====================================================

By analyzing its PES, we find a subtle quasi-linear coupling between the methyl and hydroxyl rotors in the \textit{trans} configuration (highlighted by the gray arrow in Fig.~\ref{fig:FigEthanol}-A). 
This dihedral dependence between the two functional groups is due to the electrostatic attraction between the lone pairs (negative charge) in the oxygen atom and the partially positively charged hydrogen atoms in the methyl rotor as shown in the inset in Fig.~\ref{fig:FigEthanol}-A.
Such coupling becomes clear when analyzing configurational sampling obtained from molecular dynamics simulations (Fig.~\ref{fig:FigEthanol}-B), where the dynamical implications of the coupling between the two rotors at finite temperature is evident.
Accurately capturing such interaction is crucial to correctly replicate and explain experimental measurements such as population analysis and vibrational spectra~\cite{sgdml}.

% --------------
\subsubsection{Oxygen--oxygen atom repulsion in keto--MDA}

\begin{figure}[h]
\centering
\includegraphics[width=1.0\columnwidth]{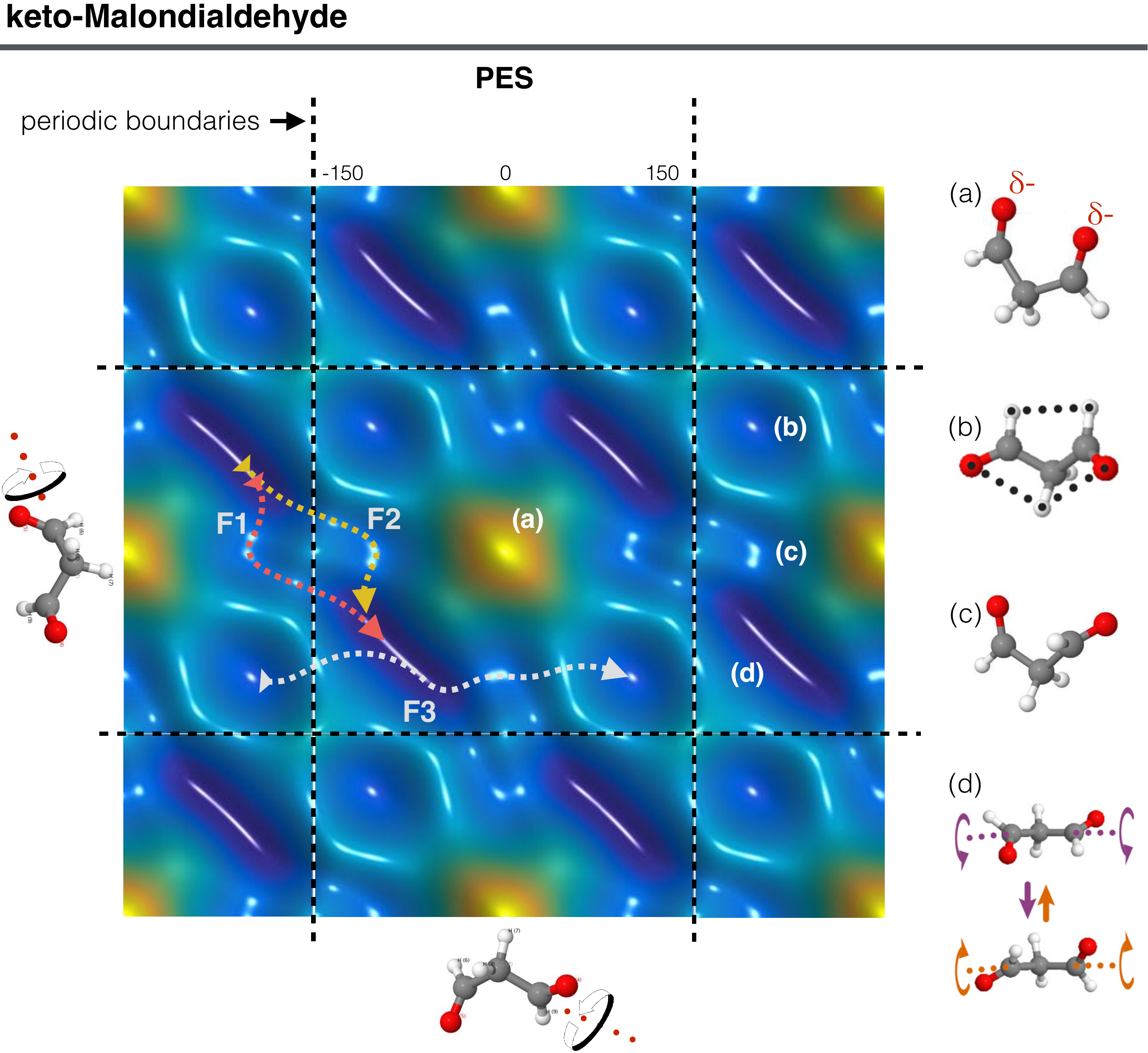}
\caption{PES for keto--MDA with periodic boundary conditions. The structure (a) leads to a steep increase in energy due to the close distance between two negatively charged oxygen atoms. (b) and (c) represent local minima and (d) display the dynamics of the global minimum. By analyzing the dynamics, the trajectories F1, F2 and F3 were found to be the most frequent transition paths.}
\label{fig:FigkMDA}
\end{figure}

From the subtle interaction described in the previous section we move to a stronger electrostatic repulsion in the keto--MDA molecule as shown in Fig.~\ref{fig:FigkMDA}. 
In a similar way as the ethanol molecule, keto-MDA can be taken as benchmark system in learning, given the complexity of its PES despite its small size (see also Fig.~\ref{fig:FigSubSampl}). 
The PES of keto--MDA molecule contains flat regions corresponding to global minimum (dark blue region in Fig.~\ref{fig:FigkMDA}) represented by the molecular structures in Fig.~\ref{fig:FigkMDA}-(d) and convoluted pathways to move from minimum to minimum (see Fig.~\ref{fig:FigkMDA}--F1, F2, F3).
Additionally, one can notice the sudden increase in the molecular energy when the two oxygen atoms are in the closest configuration as illustrated in Fig.~\ref{fig:FigkMDA}-(a).
From Fig.~\ref{fig:FigkMDA} we can see that even though the molecule only has two main degrees of freedom (the two rotors) it has a rough PES as a result of many complex interactions.
By considering the electron lone pairs in each oxygen atom and their closeness in configuration Fig.~\ref{fig:FigkMDA}-(a), it suggests that the steep increase in the energy can be primarily attributed to the electrostatic repulsion between the lone pairs in each atom.
Additionally, it could be that steric effects caused by electron cloud overlap could play also an important role.
%

% ==================== F I G U R E ===================
% =====================================================

% ==================== F I G U R E ===================
\begin{figure}[t]
\centering
\includegraphics[width=1.0\columnwidth]{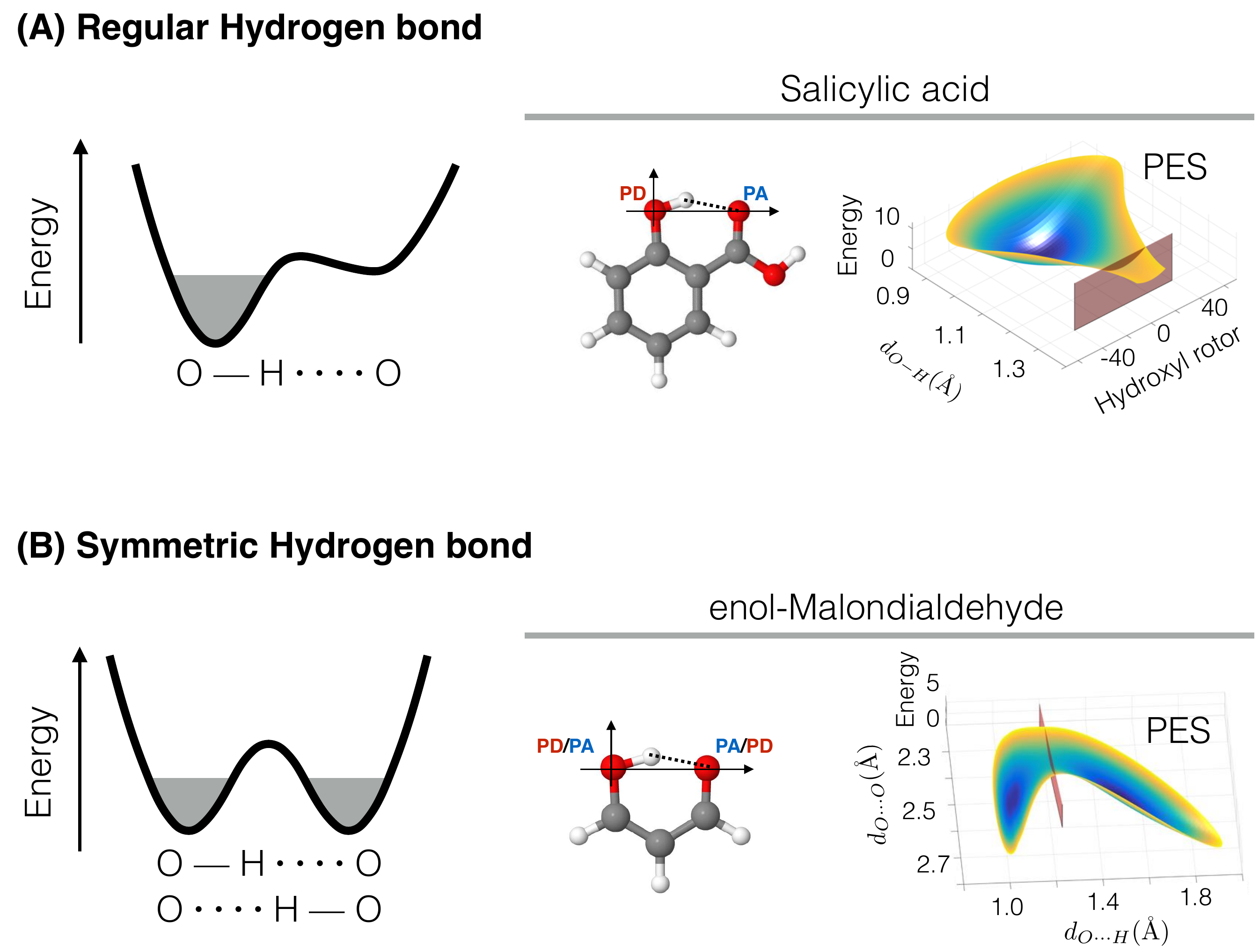}
\caption{Intramolecular hydrogen bond of (A) conventional type with salicylic acid as example and (B) symmetric type
exemplified by enol-Malondialdehyde molecule.}
\label{fig:FigHbond}
\end{figure}
% =====================================================
% --------- H--bond
\subsection{Intramolecular H--bond and proton transfer}

One of the most important phenomenon in biology and materials science is hydrogen bonding (H--bond), which is responsible of a plethora of chemical and physical effects~\cite{Scheiner2017,Hobza2002,Karpfen2009,Changwei2017,Kuhn2010}. 
Molecular mechanic force fields fail in representing this interaction due to the simple fact that we don't have an appropriate analytical model for it. 
Therefore, ML is a very promising framework to attack this problem as recently shown by the low errors accomplished by the sGDML model. 
This includes good performances in describing two different types of H--bonds: standard donor--acceptor H--bond and the symmetric H--bond. A pictorial representation of their PES and two examples of molecules containing such interaction, salicylic acid and the enol form of malondialdehyde (enol--MDA) are shown in Fig.~\ref{fig:FigHbond}.

In the particular case of regular asymmetric H--bond, as salicylic acid molecule, the interaction is a standard donor--acceptor kind of H--bond between the hydroxyl and carboxylic acid groups. 
The main characteristic of this kind of interaction consists in allowing the proton to stretch from the PD oxygen towards the PA oxygen (Fig.~\ref{fig:FigHbond}-A), which results in the well known red-shift in the stretching frequency of O--H in the participating hydroxyl group~\cite{Hobza2002,Karpfen2009,Changwei2017}.
Additionally, the H--bond also generates a blue shift in normal modes perpendicular to the H--bond, which is directly related to a O--H$\cdots$O interaction~\cite{sGDMLjcp}.
These effects can be measured experimentally via IR and Raman  spectroscopy. 

In the case of the symmetric H--bond, we observe a symmetric double-well PES as schematically represented in Fig.~\ref{fig:FigHbond}-B. 
The energetic barrier separating the two minima will be determined by the nature of the molecule under study and on the participating functional groups in the H--bond. 
In this case the symmetrized nature of the sGDML approach is crucial to consistently describe such interaction.
When the energy barrier is low, as in the case of enol-MDA: $\sim$4 kcal~$\text{mol}^{-1}$, proton transfer between the two oxygen atoms is allowed even at room temperature and it is enhanced when considering nuclear quantum effects. 
Something to highlight here is that the energy barrier can depend strongly on the level of theory used to generate the reference data. 
This is due to the intricate and subtle quantum nature of this interaction, which requires high-level quantum chemistry methods. 
It has been found that by systematically increasing the amount of electron correlation energy in the case of enol-MDA, the energy barrier decreases as $\sim$13 $\to$ $\sim$5 $\to$ $\sim$4 kcal~$\text{mol}^{-1}$ for HF $\to$ CCSD $\to$ CCSD(T), respectively~\cite{sGDMLjcp}. 
Results that demonstrate the importance of the correlation energy in such complex phenomena as the H-bond interaction and their potential effects in proton transfer. 

These two types of intramolecular H--bond are ubiquitous in nature and their presence can drastically change the chemistry and physical properties of any molecular system. Therefore the accuracy achieved by the sGDML model for the description of this interaction is particularly important.

%%%%%%%%%%%%%%%%%%%%%%%%%%%%%%%%%%%%%%%%
%======> Electronic Effects <=========== 
%%%%%%%%%%%%%%%%%%%%%%%%%%%%%%%%%%%%%%%%

% ==================== F I G U R E ===================
\begin{figure}[t]
\centering
\includegraphics[width=1.0\columnwidth]{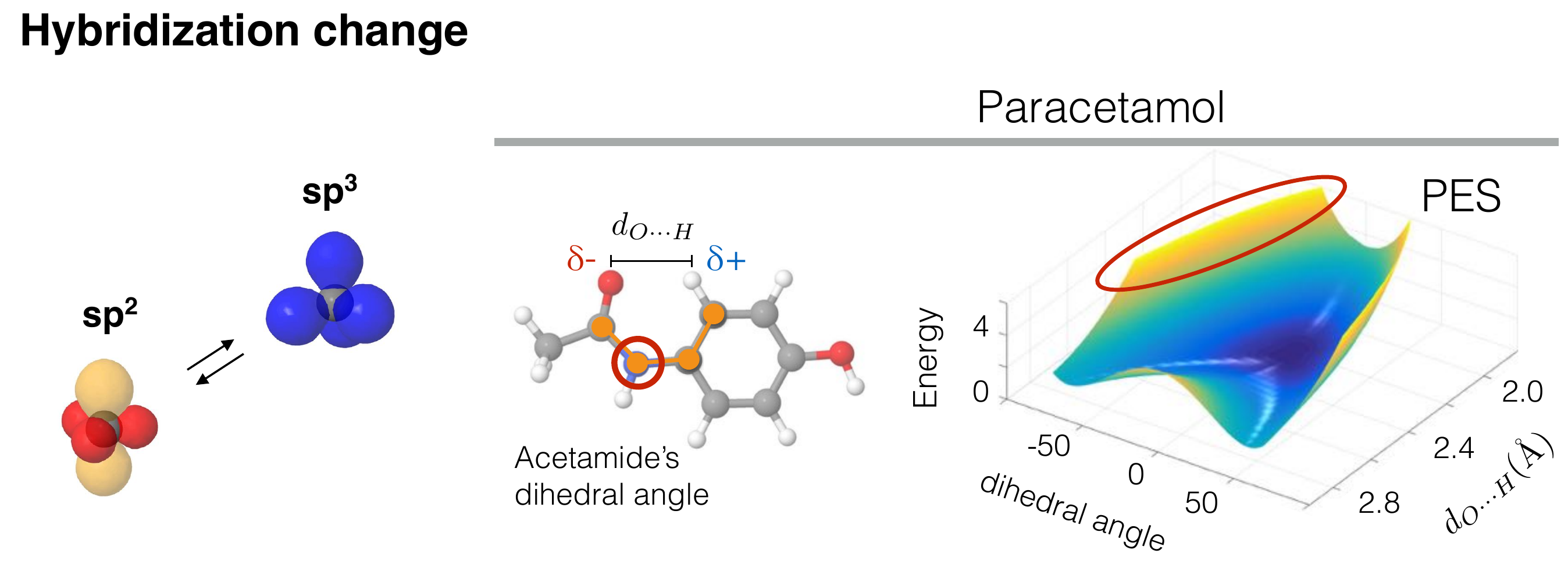}
\caption{Hybridization change. The hybridization change of the nitrogen atom is illustrated for the paracetamol molecule.}
\label{fig:FigElectroTrans}
\end{figure}
% =====================================================

\subsection{Hybridization and electronic delocalization}

Previously in this section we have shown that sGDML can learn interactions such as electrostatics and H--bonds. Here we analyze much more subtle interactions: change of hybridization and electronic interaction.
Contrary to electrostatic interactions and H--bonds, that can often be approximated and implemented in empirical FFs, purely quantum phenomena (i.e. no classical analogue) are always missing in conventional FFs.
In the case of flexible and fully data-driven model, learning  any quantum interaction coming from $-\textbf{F}=\langle \Psi^*|\partial \mathcal{H} /\partial \textbf{x}|\Psi \rangle$, is a trivial task accomplished without relying on prior knowledge of the phenomena or its connection to any classical electrodynamic or mechanical concepts.
Two examples are capturing the configuration and energetic features associated to changes in hybridization states and $n\to\pi^*$ interactions.

In general, changes in the molecular electronic state is related to rearrangements of the electronic configuration to minimize the energy for a given molecular conformation. 
This could be for example a transition from a singlet to a triplet state, as in the case of some metallic clusters or molecules~\cite{azobznTriplet2004}, or locally modifying an atomic hybridization state.
The paracetamol molecule, for example, is a system that presents a $sp^2 \leftrightarrow sp^3$ hybridization change in the nitrogen atom for configurations in which the dihedral angle of the acetamide group is increasing while keeping the interatomic distance $d_{O\cdots H}$ constant (see Fig.~\ref{fig:FigElectroTrans}).
This generates a steep energy increase as illustrated by yellow regions in the PES in Fig.~\ref{fig:FigElectroTrans}. 
In that region of configuration space the conjugated state in the molecule breaks, given that the nitrogen atom changes its hybridization state from $sp^2 \to sp^3$.

Another important, but less studied electronic interaction, is the overlap between occupied (lone pair $n$) and antibonding ($\pi^*$) orbitals: $n\to\pi^*$ interactions. 
The analysis of this interaction is beyond the scope of the current book chapter but it is worth to mention a couple of things.
The $n\to\pi^*$ interactions is a ubiquitous interaction in biological and other molecular systems but only recently it was found the importance of such weak interaction~\cite{AceProNMe2016}. 
In particular, it plays a very decisive role in the dynamics of the aspirin molecule. 
The $n\to\pi^*$ attraction interaction is responsible for the binding between the ester and carbonyl groups, defining the structure of the global minimum even at room temperature~\cite{sgdml}. 

There are many other electronic effects (e.g. hyperconjugation, configuration dependent charge densities, Jahn--Teller effect, $\pi$--hole interactions, etc.~\cite{Exp03Spectro2017}) for which we don't have analytical approximations. Therefore, they can not be incorporated in conventional FFs limiting their reliability and predictive power.
The rigorous requirement of accurately capturing such effects in ML models is justified by the increasing precision in state-of-the-art spectroscopic experimental results~\cite{Fielicke2008,Fielicke2012,Exp01Spectro2010,Exp01Spectro2016,Exp01Spectro2017,Exp02Spectro2017,Exp03Spectro2017} which demand computationally inexpensive and highly accurate PESs to interpret and obtain further insights.  

In summary, in this section we have analyzed a wide variety of interatomic interaction via high fidelity energy landscapes learning with the sGDML model. In particular, we described hydrogen bonds, electrostatic and electronic interactions.
But as a final comment in this chapter, it is fair to ask the question: How relevant these interactions are in larger systems?
This is because up to this moment, we have shown the importance of these phenomena in small molecules where it is understandable that such interactions play a major role. 
The answer is yes, all these interactions together play a major role in protein folding as recently suggested by Deepak \textit{et al.}~\cite{AceProNMe2016}. The edge where some proteins fold is a result of a complex interplay between many of the interactions analyzed here.
Consequently, one of the main challenges in the route to model bigger systems is to preserve the reliability of the sGDML framework on describing such interactions. 

\section{Conclusions}\label{sec:concl}
% -- summary -- 
In this book chapter we have presented the construction of molecular force fields using the symmetrized Gradient Domain Machine Learning model. 
In particular, we have introduced what are the desirable requirements of machine learning force fields from the point of view of physics and computation efficiency. 
In this context, the sGDML framework is able to reconstructs high-dimensional manifolds embedded in the training data even from a few 100s of samples. 
Achievement that allows the use of highly-accurate \textit{ab initio} reference data such as the ``gold standard" CCSD(T) method. 
The flexibility of such universal approximator comes from its fully data-driven nature, characteristic that grants the adaptability to describe any quantum interaction coming from $-\textbf{F}=\langle \Psi^*|\partial \mathcal{H} /\partial \textbf{x}|\Psi \rangle$.
Here we have also described a simple way to systematically increase the level of theory from DFT to CCSD(T) by the subsampling--and--recomputing method, keeping in mind that the DFT's PES is already close to the CCSD(T) one. 

% -- Advantages --
The main advantages over other machine learning methods are: 
\textit{(i)} highly data efficient originated by being trained in the gradient domain, 
\textit{(ii)} its robustness acquired by modeling all atomic interactions globally without any inherent non-unique partitioning of the energy or force, 
\textit{(iii)} it encodes the fundamental physical law of energy conservation as well as 
\textit{(iv)} atomic indistinguishability as a prior, correctly representing spatial and temporal symmetries.

Some challenges remain to be solved within the sGDML framework, mainly consisting in how to extend its applicability to larger systems. 
Many of the advantages of the model are related to its global nature, unfortunately this also imposes limits on the maximum size of the molecules that can be considered as well as the training set size. 
Solving this fundamental problem requires careful and  well-reasoned fragmentation schemes to divide the problem into smaller independent subproblems without compromising its robustness. 
A possible direction to go can be a data-driven approach in a way that is tailored to preserving the intricate phenomena and quantum interactions studied in this chapter. 
The existence of such approach would benefit from the explicit knowledge of fluxional symmetries within the system and well defined functional groups.
In its current formulation, the sGDML framework captures different types of interaction as well as interaction scales, with no need to separating them. 
Nevertheless, an explicit decoupling of long-range interactions, e.g. van der Waals forces, could be a new avenue to further increase its applicability to increasingly larger and complex molecules. 

\section{Acknowledgments}
S.C., A.T., and K.-R.M. thank the Deutsche Forschungsgemeinschaft, Germany (projects MU 987/20-1 and EXC 2046/1 [ID: 390685689]) for funding this work. A.T. is funded by the European Research Council with ERC-CoG grant BeStMo. This work was supported by the German Ministry for Education and Research as Berlin Big Data Centre (01IS14013A) and Berlin Center for Machine Learning (01IS18037I). This work was also supported by the Information \& Communications Technology Planning \& Evaluation (IITP) grant funded by the Korea government (No. 2017-0-00451). This publication only reflects the authors views. Funding agencies are not liable for any use that may be made of the information contained herein. Part of this research was performed while the authors were visiting the Institute for Pure and Applied Mathematics, which is supported by the National Science Foundation, United States .


%\bibliography{referenc}% Produces the bibliography via BibTeX.

\end{document}